\documentclass[a4paper,11pt]{article}
\usepackage[utf8x]{inputenc}
\usepackage[margin=0.9in]{geometry}
\usepackage{amsmath}
\usepackage{latexsym}
\usepackage{amssymb}
\usepackage[english]{babel}
\usepackage{color}
\usepackage[multiple]{footmisc}
\usepackage{jheppub}
\pdfoutput=1
\usepackage{breakurl}
\usepackage{mathrsfs}
\usepackage{amsfonts}
\usepackage{mathrsfs}
\usepackage{slashed}
\usepackage{epigraph}
\usepackage{pifont}
\usepackage{fancyhdr}
\usepackage{graphicx}
\usepackage{longtable}
\usepackage{makecell}
\usepackage{amscd}
\usepackage{tabu}
\usepackage{bm}
\usepackage{enumerate}
\usepackage{caption}
\usepackage{subcaption}
\usepackage{bbold}
\usepackage{tikz}
\usepackage[compat=1.1.0]{tikz-feynman}
\usepackage{bbm}
\usepackage{placeins}
\usepackage{hyperref} 
\usepackage{cleveref}
\hypersetup{linktocpage=true}
\usepackage[all]{hypcap}
\hypersetup{
   bookmarks=true,         
   unicode=false,          
   pdftoolbar=true,        
   pdfmenubar=true,        
   pdffitwindow=false,     
   pdfstartview={FitH},    
   pdftitle={My title},    
   pdfauthor={Author},     
   pdfsubject={Subject},   
   pdfcreator={Creator},   
   pdfproducer={Producer}, 
   pdfkeywords={keyword1} {key2} {key3}, 
   pdfnewwindow=true,      
   colorlinks=true,        
   linkcolor=blue,         
   citecolor=magenta,      
   filecolor=magenta,      
   urlcolor=cyan           
}

\usepackage{fancyhdr}
\usepackage{booktabs}
\usepackage{tikz}
\usetikzlibrary{decorations}
\usetikzlibrary{arrows,intersections,shapes,decorations.markings}
\usetikzlibrary{arrows.meta}
\usetikzlibrary{shapes.misc}
\usetikzlibrary{decorations.pathmorphing}

\preprint{MITP/20-068,P3H-20-069,TTK-20-41
}
\title{Collider constraints on dark mediators}
\author{Hanna Mies$^{* a}$}
\author{Christiane Scherb$^{\dagger b}$}
\author{and Pedro Schwaller$^{\dagger c}$}

\affiliation{\normalfont{$^*$Institute for Theoretical Particle Physics and Cosmology, RWTH Aachen University, 52056 Aachen, Germany}}
\affiliation{\normalfont{$^\dagger$PRISMA$^+$ Cluster of Excellence \& Mainz Institute for Theoretical Physics, Johannes Gutenberg University, 55099 Mainz, Germany}}
\emailAdd{$^a$mies@physik.rwth-aachen.de}
\emailAdd{$^b$cscherb@uni-mainz.de}
\emailAdd{$^c$pedro.schwaller@uni-mainz.de}

\abstract{\\
We explore the constraints current collider searches place on a QCD-like dark sector. A combination of multi-jet, multi-jet plus missing energy and emerging jets searches is used to derive constraints on the mediator mass across the full range of the dark meson lifetimes for the first time. 
\\The dark sector inherits a flavour structure from the coupling between the dark quarks and the SM quarks through the mediator. When this is taken into account, the differently flavoured dark pions become distinguishable through their lifetime. We show that also in these cases the above mentioned searches remain sensitive, and we obtain limits on the mediator mass also for the flavoured scenario. 
\\
We then contrast the constraints from collider searches with direct detection bounds on the dark matter candidate itself in both the flavoured and unflavoured scenario. Using a simple prescription it becomes possible to display all constraints in the dark matter and mediator mass plane. Constraints from direct detection tend to be stronger than the collider constraints, unless the coupling to the first generation quarks is suppressed, in which case the collider searches place the most stringent limits on the parameter space. 
}

\begin{document}

\maketitle

\section{Introduction}

The quest for dark matter is one of the most pressing open problems in particle physics, and one of the main drivers for searches for new physics at the LHC experiments. In order to make sure that "no stone is left unturned", all potential signatures of dark matter models have to be identified and searched for. In this context, it is crucial to understand when a new search strategy has to be implemented and when a new signal is already probed by existing search channels. 
Furthermore, it is important to understand the complementarity of collider probes with the sensitivity of other searches such as direct detection of dark matter, or constraints coming from cosmology, flavour data or fixed target experiments. 

In this work, we answer these questions for scenarios where a t-channel mediator communicates between the standard model (SM) and a QCD-like dark sector. This complements and extends similar studies for minimal dark sectors~\cite{Papucci:2014iwa,Abdallah:2014hon,Abdallah:2015ter}, as well as searches for complex dark sectors coupled via s-channel mediators~\cite{Cohen:2015toa,Cohen:2017pzm,Pierce:2017taw,Park:2017rfb,Cheng:2019yai,Bernreuther:2019pfb,Cohen:2020afv,Bernreuther:2020vhm,Kar:2020bws,Bernreuther:2020xus}. 

At colliders, QCD-like dark sectors with a sufficiently low confinement scale can produce spectacular signals. Once dark quarks are produced, they undergo a parton shower and hadronisation process, resulting in jets of hidden sector mesons and baryons~\cite{Strassler:2006im,Han:2007ae,Strassler:2008fv,Baumgart:2009tn,Beauchesne:2018myj,Kribs:2018ilo,Beauchesne:2017yhh,EmergingJets}. Depending on the mediator structure, some or all dark hadrons eventually decay back to SM particles, possibly with a decay length in the millimeter to meter range. For purely hadronic decays of dark mesons, a search strategy for such "emerging jets" was proposed in~\cite{EmergingJets}, which relies on the fact that more and more parts of the jet appear in the detector as one moves away from the interaction point. The first search for emerging jets was recently published by the CMS Collaboration~\cite{jetEmergingJet1} and was able to put constraints on mediator masses up to 1.5~TeV for lifetimes between 1~mm and 1~m. For a survey of other long lived particle searches currently being discussed, see the review~\cite{Alimena:2019zri}. 

Here, we derive for the first time constraints on the mediator mass in the model introduced in~\cite{EmergingJets} and summarised in~\cref{sec:model} for the full range of dark meson lifetimes, by combining the emerging jets search with constraints from multi-jet and multi-jet plus missing energy channels (c.f.~\cref{sec:recast1}). Furthermore, we perform a recast of all search channels for scenarios with more than one characteristic dark meson lifetime, so called flavoured emerging jets, and show {in~\cref{sec:recast2} that also in these cases strong constraints on the mediator mass exist. 

Regarding the complementarity, we developed a method to display the collider constraints on this model in the usual mediator mass - dark matter mass plane and combine them with direct detection bounds. We also include the collider bounds in updated constraints on the flavour structure of the model, which remains testable in a variety of experiments such as Belle II \cite{Aushev:2010bq,Buras:2014fpa,Altmannshofer:2009ma,Kou:2018nap}, NA62 \cite{NA62:2017rwk}, SHiP \cite{Alekhin:2015byh}, as well as DM direct detection experiments (c.f.~\cref{sec:constraints}). To evade the very strong bounds from the XENON1T experiment~\cite{Xenon}, it can be necessary to suppress the coupling of the mediator to the first generation of SM quarks, which leads us to the concept of a \textit{strange dark sector}. We conclude with a discussion of some new ideas for combining different LLP search strategies to further improve the sensitivity for our and similar dark sectors in~\cref{sec:summary}.

\section{The model}
\label{sec:model}
\subsection{Model description}

We start with a brief summary of the model that was introduced in~\cite{EmergingJets,FlavouredDarkSec}. There, the standard model (SM) gauge group is extended by a new $SU(N_D)$ group under which all SM particles are neutral. Furthermore, $n_D$ Dirac fermions or \textit{dark quarks} $Q_D$ are introduced which transform in the fundamental representation of $SU(N_D)$, and which are neutral with respect to the SM gauge interactions. Finally, a scalar $X_D$, charged under both QCD and the \textit{dark QCD} $SU(N_D)$, allows for Yukawa-type interactions between quarks and dark quarks, and thus serves as the mediator between the visible and dark sector. 

In the UV, the Lagrangian of the model is given by
\begin{equation}
    \mathcal{L}\supseteq \mathcal{L}_{SM} -\frac{1}{4} G_{D,\mu\nu}G_D^{\mu\nu}+ i\bar{Q}_D \slashed D Q_D - m_{Q_D} \bar{Q}_D Q_D + D_\mu X^\dagger_D D^\mu X_D+ m_{X} |X_D|^2 + \mathcal{L}_{Yuk}, \label{eqn:lagrange1}
\end{equation}
where $G_D$ is the field strength tensor of the $SU(N_D)$ dark gluons, $m_{Q_D}$ is the Dirac mass which respects the dark flavour symmetry, and $D_\mu$ and $\slashed D$ are the appropriate covariant derivatives. 
The coupling to the SM arises through a Yukawa interaction of the form
\begin{equation}
    \mathcal{L}_{Yuk}= -\kappa_{ij} \bar{d}_{Rj} Q_{D_{Li}} X_D + h.c.,
    \label{LagrangianYukawa}
\end{equation}
which is allowed if $X_D$ transforms as (3,$\bar{N}_D$) under $SU(3)_{colour}$ x $SU(N_D)_D$ and has a hypercharge $Y=-\frac{1}{3}$. Different charge assignments which instead would allow couplings to up-type quarks, quark doublets or to leptons will be discussed elsewhere. 

We assume that $SU(N_D)$ is asymptotically free and confines at a scale $\Lambda_D$, and that the dark flavour symmetry is only weakly broken by the dark quark mass term. In that case at scales below $\Lambda_D$ the model contains $n_d^2-1$ pseudo Nambu-Goldstone bosons, the \textit{dark pions} $\pi_D$, with masses $m_{\pi_D} \ll \Lambda_D$. For all practical purposes we will take $N_D=3$ and  $n_d \in \{2, 3\}$ in the remainder of this work.

The mediator $X_D$ can be efficiently pair-produced at hadron colliders thanks to its colour charge. It will then undergo a prompt decay to a quark and a dark quark, leading to one QCD jet and one dark jet per mediator. Our goal in this work is to obtain robust constraints on the mediator mass $m_X$ and on the dark pion mass $m_{\pi_D}$, and we will therefore model the potentially complicated dark sector dynamics with as few parameters as possible.

\subsection{Phenomenological parameters}
In the UV, the model is specified by the masses of the dark quarks $m_{Q_D}$ and the mediator $m_{X}$, the Yukawa couplings $\kappa_{ij}$, and the confinement scale $\Lambda_D$ that determines the dark strong coupling. We will first consider the case of universal dark Yukawa couplings, $\kappa_{ij} = \kappa_0$. Then the dark pion lifetime is given by 
\begin{equation}
    c\tau_{\pi_D}= 80mm \cdot \frac{1}{\kappa_0^4}\cdot \biggl(\frac{2~\rm GeV}{f_{\pi_D}}\biggr)^2 \biggl(\frac{100~\rm MeV}{m_d}\biggr)^2 \biggl(\frac{2~\rm GeV}{m_{\pi_D}}\biggr) \biggl(\frac{m_{X}}{1000~\rm GeV}\biggr)^4,
    \label{lifetime}
\end{equation}
where $m_d$ is the mass of the heaviest down-type quark that is kinematically accessible. Here, two new phenomenological parameters appear, the dark pion mass $m_{\pi_D}$ and the dark pion decay constant $f_{\pi_D}$. 

Since these parameters are not analytically calculable from the UV theory, we have to choose them based on our experience with QCD at low energies. The dark pion mass is proportional to the dark quark mass which explicitly breaks the chiral symmetry, so we can instead take $m_{\pi_D} \ll \Lambda_D$ as a free parameter. Furthermore, in QCD we have $f_\pi \approx m_\pi$, so we choose $f_{\pi_D} = m_{\pi_D}$ to reduce the number of independent free parameters\footnote{Naive dimensional analysis would suggest that $f_{\pi_D} \approx \Lambda_D/(4 \pi)$, i.e. independent of the dark quark mass. In practice we always vary $m_{\pi_D}$ such that this relation is not grossly violated, so it is not necessary to treat $f_{\pi_D}$ as an independent parameter.}. Finally, we can trade the coupling $\kappa_0$ for the lifetime $\tau_{\pi_D}$. The phenomenology of the model will therefore be mainly controlled by three parameters:  
\begin{align}
	m_{\pi_D}\,,\qquad \tau_{\pi_D}\,,\qquad m_{X}\,,
\end{align}
which we vary independently. 

Going beyond the case of universal Yukawa couplings $\kappa$, the number of free parameters grows drastically. Yet in practice the main effect is that the different dark pions become distinguishable by their lifetime. While up to 5 different lifetimes can appear\footnote{For general $n_d$ there are $n_d^2-1$ Goldstone bosons. In the absence of CP violation, the off-diagonal pions form particle-anti particle pairs, therefore one expects $n_d(n_d+1)/2-1$ individual lifetimes.}, in~\cref{sec:recast2} we limit the modelling of this effect by allowing two independent lifetimes, i.e. 
\begin{align}
	m_{\pi_D}\,,\qquad \tau_{\pi_{D1}}\,,\qquad \tau_{\pi_{D2}}\,,\qquad m_{X}\,,
\end{align}
with varying relative probabilities. This will give us a good idea of how the sensitivities of the individual searches degrade, while keeping the complexity of the model at a manageable level.

\subsection{Implementation}
\label{sec:implementation}
Different tools have been used to implement the model, generate events and then decay and hadronise the particles in these events. 

To generate parton level events, UFO files for the model were produced using the~\textsc{FeynRules}~\cite{FeynRules} package. Here, only the mediator $X_D$ and the dark quarks $Q_D$ and a universal coupling $\kappa_0$ were implemented. The resulting UFO file is read into \textsc{MadGraph5} version 5.1.6 \cite{MadGraph} to create pair produced mediators $X_D$ in the process $p p \to X_D X^\dagger_D$.
Compared with directly generating events in \textsc{Pythia8}, the advantage of this setup is that additional processes can now be included. In particular we find that the $t$-channel exchange of $Q_D$ (see~\cref{FeynmanDiagrams1}) contributes significantly for $\kappa_0 \gtrsim 0.1$, even leading to a small suppression of the cross section due to destructive interference around $\kappa_0 \approx 0.5$, as shown in~\cref{kappa}. Other processes which can now be generated include the radiation of additional hard jets and the single production of a mediator together with a dark quark from a quark-gluon initial state.

The generated events are read into \textsc{Pythia8} version 8.2.3.5 \cite{Pythia}. Using the Hidden Valley (HV) model implementation \cite{HVpythia,Carloni:2010tw,Carloni:2011kk}, the heavy particles in the event are decayed, showered, and hadronised both in QCD and in the dark sector. Finally, the dark pions can also be decayed back to SM particles, while it is assumed that all heavier particles in the dark QCD sector are either stable (missing energy) or promptly decay to dark pions. 

Additional phenomenological parameters appear in the HV model implementation in \textsc{Pythia8}. As done in~\cite{EmergingJets}, we use the relations $2 m_{\pi_D} = m_{Q_D} = \Lambda_D  = 1/2 m_{\rho_D}$, since this gives the best agreement of the shower multiplicity with the theoretical prediction. Here, the dark quark mass should not be considered to be the same as the one appearing in~\cref{eqn:lagrange1}, but as a phenomenological "constituent" quark mass.

For the flavoured scenario, which will be considered in~\cref{sec:recast2}, individual lifetimes for the different dark pions should be set. However,
\textsc{Pythia8} can only distinguish up to two different dark pions, therefore implementing a fully flavoured dark sector is not possible. As discussed above, we can however simulate scenarios with two different characteristic dark pion lifetimes, and thereby capture the main effect that flavour will have on the phenomenology of the model. 
The relative abundance of the two lifetimes is indirectly controlled by the number of dark flavours, $n_d$. While the the ratio of diagonal to off-diagonal degrees of freedom is $(n_d-1)/(n_d^2-1-(n_d-1)) = 1/n_d$, the string fragmentation model implemented in \textsc{Pythia8} actually gives a ratio of $1/(n_d-1)$ in generated events.

\section{Recast of LHC searches}
\label{sec:recast1}
At hadron colliders dark mediators are produced via single or pair production. Here, we focus on pair production. The respective Feynman graphs are shown in~\cref{FeynmanDiagrams1}. Each mediator decays to a quark and a dark quark, which result in jets composed of hadrons and dark hadrons, respectively. 
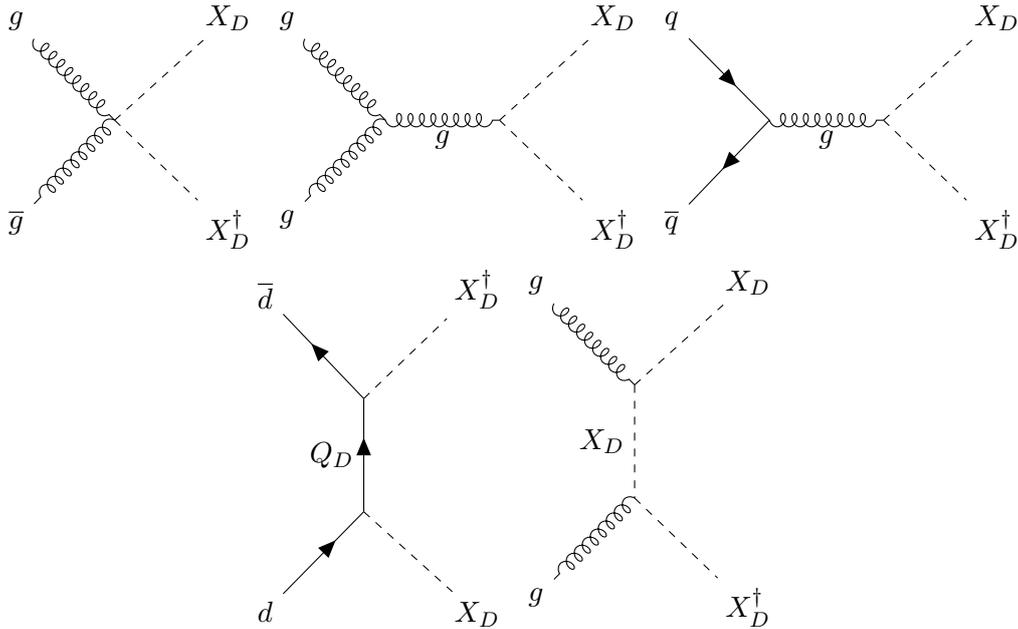
\begin{figure}[t]
	\begin{center}
		\begin{tikzpicture}
		\begin{feynman}
		\vertex (b);
		\vertex [above left=of b] (f1) {\(g\)};
		\vertex [below left=of b] (f2) {\(\overline{g}\)};
		\vertex [above right=of b] (f3) {\(X_D\)};
		\vertex [below right=of b] (f4) {\(X^\dagger_D\)};
		
		\diagram* {
			(f1) -- [gluon] (b) -- [gluon] (f2),
			(b) -- [scalar] (f3),
			(b) -- [scalar] (f4),};
		\end{feynman}
		\end{tikzpicture}
		\begin{tikzpicture}
		\begin{feynman}
		\vertex (b);
		\vertex [above left=of b] (f1) {\(g\)};
		\vertex [below left=of b] (f2) {\(g\)};
		\vertex [right=of b] (c);
		\vertex [above right=of c] (f3) {\(X_D\)};
		\vertex [below right=of c] (f4) {\(X^\dagger_D\)};
		
		\diagram* {
			(f1) -- [gluon] (b) -- [gluon] (f2),
			(b) -- [gluon, edge label'=\(g\)] (c),
			(c) -- [scalar] (f3),
			(c) -- [scalar] (f4),};
		\end{feynman}
		\end{tikzpicture}
		\begin{tikzpicture}
		\begin{feynman}
		\vertex (b);
		\vertex [above left=of b] (f1) {\(q\)};
		\vertex [below left=of b] (f2) {\(\overline{q}\)};
		\vertex [right=of b] (c);
		\vertex [above right=of c] (f3) {\(X_D\)};
		\vertex [below right=of c] (f4) {\(X^\dagger_D\)};
		
		\diagram* {
			(f1) -- [fermion] (b) -- [fermion] (f2),
			(b) -- [gluon, edge label'=\(g\)] (c),
			(c) -- [scalar] (f3),
			(c) -- [scalar] (f4),};
		\end{feynman}
		\end{tikzpicture}
		
		\begin{tikzpicture}
		\begin{feynman}
		\vertex (b);
		\vertex [above left=of b] (f1) {\(\overline{d}\)};
		\vertex [above right=of b] (f2) {\(X^\dagger_D\)};
		\vertex [below=of b] (c);
		\vertex [below left=of c] (f3) {\(d\)};
		\vertex [below right=of c] (f4) {\(X_D\)};
		
		\diagram* {
			(f1) -- [anti fermion] (b) -- [scalar] (f2),
			(b) -- [anti fermion, edge label'=\(Q_D\)] (c),
			(c) -- [anti fermion] (f3),
			(c) -- [scalar] (f4),};
		\end{feynman}
		\end{tikzpicture}
		\begin{tikzpicture}
		\begin{feynman}
		\vertex (b);
		\vertex [above left=of b] (f1) {\(g\)};
		\vertex [above right=of b] (f2) {\(X_D\)};
		\vertex [below=of b] (c);
		\vertex [below left=of c] (f3) {\(g\)};
		\vertex [below right=of c] (f4) {\(X^\dagger_D\)};
		
		\diagram* {
			(f1) -- [gluon] (b) -- [scalar] (f2),
			(b) -- [scalar, edge label'=\(X_D\)] (c),
			(c) -- [gluon] (f3),
			(c) -- [scalar] (f4),};
		\end{feynman}
		\end{tikzpicture}
	\end{center}
	\caption{Feynman diagrams for the pair production of the mediator.}
	\label{FeynmanDiagrams1}
\end{figure}

 \begin{figure}[t]
 	\centering
	\includegraphics[width=0.47\textwidth]{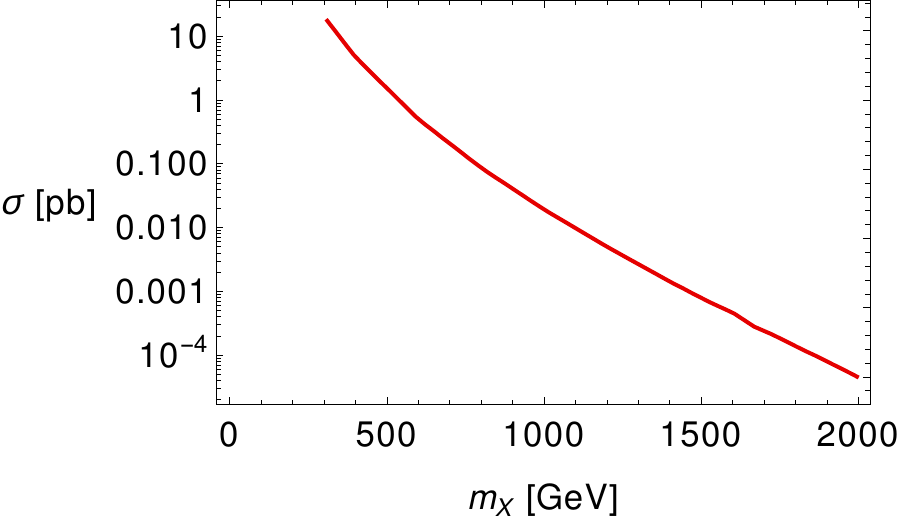}
	\includegraphics[width=0.47\textwidth]{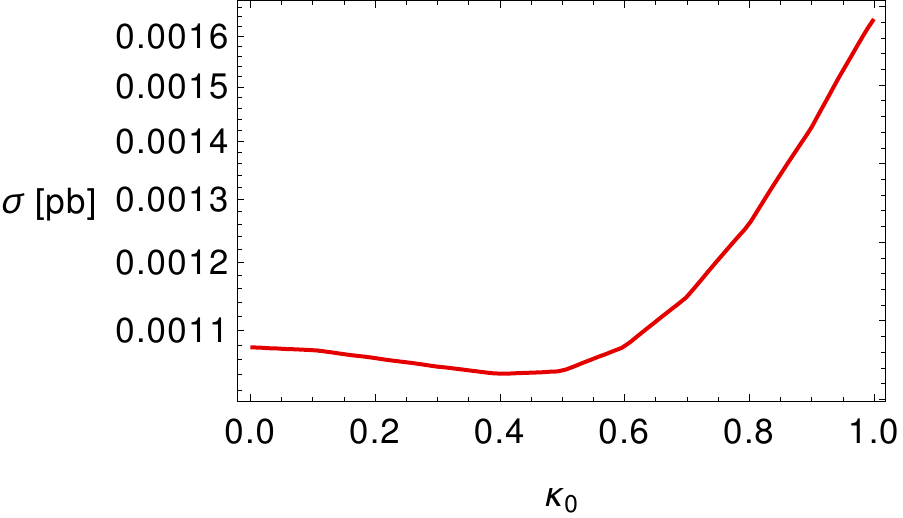}

 \caption{Mediator pair production cross section, as a function of the mediator mass $m_X$ for fixed $\kappa_0=1$(left) and as a function of $\kappa_0$ for $m_X = 1200$~GeV (right). }
 \label{kappa}
 \end{figure}

The pair production cross section at the LHC is shown in~\cref{kappa} as a function of the mediator mass $m_X$ and of the Yukawa coupling $\kappa$. As can be seen in the right figure, for order one Yukawa couplings, the contribution from $t$-channel exchange of dark quarks can increase the cross section by up to 50\%, an effect that was missed previously in ~\cite{EmergingJets}. There is also a small region where destructive interference reduces the signal rate.

The produced heavy dark hadrons decay promptly into dark pions. Depending on the lifetime of the dark pions three different final states can be considered: (1) if the dark pions are long lived ($c\tau_{\pi_D}\geq0.05$~m) 
 they appear as missing energy in a detector, (2) the dark pions have intermediate lifetimes ($0.001$~m$\leq c\tau_{\pi_D} \leq 1$~m) and decay to emerging jets \cite{EmergingJets,jetEmergingJet1} or (3) the dark pions decay prompt ($c\tau_{\pi_D} \leq 0.1$~m), 
 see~\cref{decayranges}. 
 
 We therefore expect that different searches provide the best sensitivity to the model for different values of the dark pion lifetime. Our goal in the following is to show that this is indeed the case, and that by combining the different search strategies, the full range of dark pion lifetimes can be covered. Since we always get at least two QCD jets from the SM quarks, from long to short lifetimes the signatures we expect are: (1) two jets and missing energy, (2) two jets and two emerging jets and (3) four prompt jets.

\begin{figure}[t]
\centering
\includegraphics[width=0.6\textwidth]{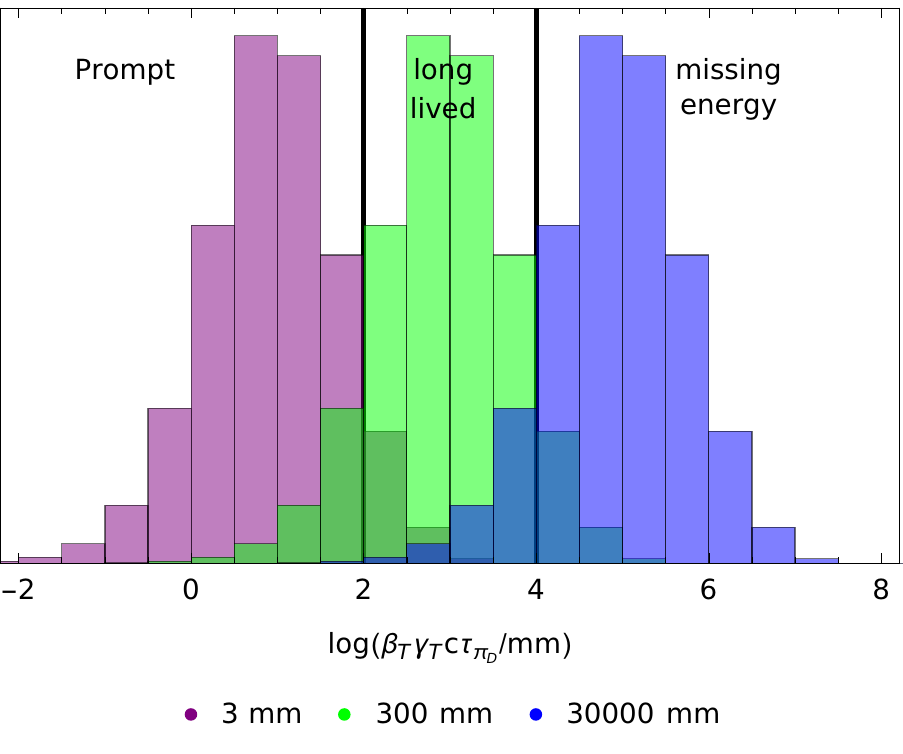}
\caption{Radial distribution of the dark pion decay lengths in the laboratory frame, for various values of $c\tau_{\pi_D}$, $m_X= 1200$~GeV and $m_{\pi_D} = 2$ GeV. The black lines indicate the typical extent of the LHC detectors, i.e. where displaced decay signatures can in principle be detected, while earlier (later) decays will appear as prompt (missing energy) signatures. }
\label{decayranges}
\end{figure} 

\FloatBarrier
\subsection{Jets plus MET search and four jets search}
To find the allowed parameter space for long-lived dark pions (Jets+MET search)   and  dark pions that decay promptly (four jets search) we recast the LHC searches in \cite{MetNew} and in \cite{4jets}, respectively.  The requirements that need to be fulfilled for an event to count as signal are summarised in~\cref{tab:requirements} for both searches.
\begin{table}[t]
	\centering
	\begin{tabular}{|c|c|}
		\hline
		$\boldsymbol{jets+MET\text{ }search}$&$\boldsymbol{4\text{ }jets\text{ }search}$\\
		\hline
		$N_{jet} \geq 2$ for jets with $\eta\leq 2.4$    & $N_{jet}\geq4$ for jets with $|\eta|\leq 2.4$           \\ \hline
		$H_T>300$~GeV, & $p_T > 120$ GeV for each jet          \\ 
		where $H_T$ is the scalar $p_T$ sum of the jets&\\\hline
		$|\Vec{H}_T^{miss}| > 300 $ GeV,  $p_T$ sum with $|\eta|<5$ & at least two tracks with $p_T>400$ MeV         \\ 
		with $\Vec{H}_T^{miss}$ the negative of the vector&\\\hline
	\end{tabular}
	\caption{Summary of the requirements for an event to count as signal for the jets plus MET search and the four jets search.}
	\label{tab:requirements}
\end{table} 

In case of the jets plus MET search for each simulated event the point of the dark pion's decay to a visible particle has to be recorded and checked if it decays inside or outside the hadronic calorimeters (HCAL). For dark pions decaying outside the HCAL momenta in x- and y-directions are summed to obtain $\Vec{p}_T^{miss}$. Summing $\Vec{p}_T^{miss}$ over all dark pions of an event gives the value of the MET. Hereby, all particles with $|\eta|<5$ are considered to get a better accuracy of the MET. If the total MET of an event exceeds $300$ GeV and the additional requirements in~\cref{tab:requirements} are fulfilled it is counted as signal. 

For the four jets search the  \textsc{Pythia8} variables \textit{ParticleDecays:xyMax} and \textit{ParticleDecays:limitCylinder} are used to limit the particle decays to the interior of the detector. All particles that would have decayed outside the detector now remain stable. Thus, it is easy to distinguish dark pions that are visible to the detector from dark pions that are invisible.
Jets, formed by all visible particles, are created with the help of the program SlowJet \cite{SlowJet}.
Thereby, the cone size is set to $R = 0.4$, the anti-$k_T$ jet algorithm is used and the pseudo-rapidity is limited to $|\eta| < 2.4$. If the resulting event fulfils all the requirement of the four jets search in~\cref{tab:requirements} it is counted as signal.

The above described procedures have been performed for different lifetimes ($0.05$~m $\leq c\tau \leq 100$~m for the jets plus MET search and $0.001$~mm $\leq c\tau\leq 100$~mm for the four jets search) of the dark pions with $10\,000$ events per lifetime to find how the signal acceptance depends on the lifetime. We denote the number of events that pass the required cuts $N(c\tau_{\pi_D})$ with $N(c\tau_{\pi_D}) \leq 10\,000$. 

We set a baseline point for both the jets plus MET search and the four jet search to which we compare all other considered lifetimes to obtain an acceptance rate: For very large lifetimes ($c\tau_{\pi_D}\geq 100$ m) the dark pions can be seen as stable at detector scales. For the jets plus MET search this "infinite" lifetime is used as the baseline. For smaller lifetimes only a fraction of the dark pions decays outside the detector. 
Analogously, in the four jets search the smallest lifetime gives the largest acceptance and is considered to be a nearly prompt decay. In our case, this is the number of events for $c\tau =0.001$~mm, which we call $N_{jets}(0.001$~mm$)$ and take as our baseline. We calculate the acceptance rate for the jets plus MET search as $N_{\rm MET}(c\tau_{\pi_D})/N_{\rm MET}(100$~ m). Similarly, for the four jets search the acceptance is $N_{jets}(c\tau_{\pi_D})/N_{jets}(0.001$~mm). The results are shown in~\cref{fig:acceptance}.
\begin{figure}[t]
    \centering
    \begin{subfigure}{0.47\textwidth}
    	\centering
        \includegraphics[width=\linewidth]{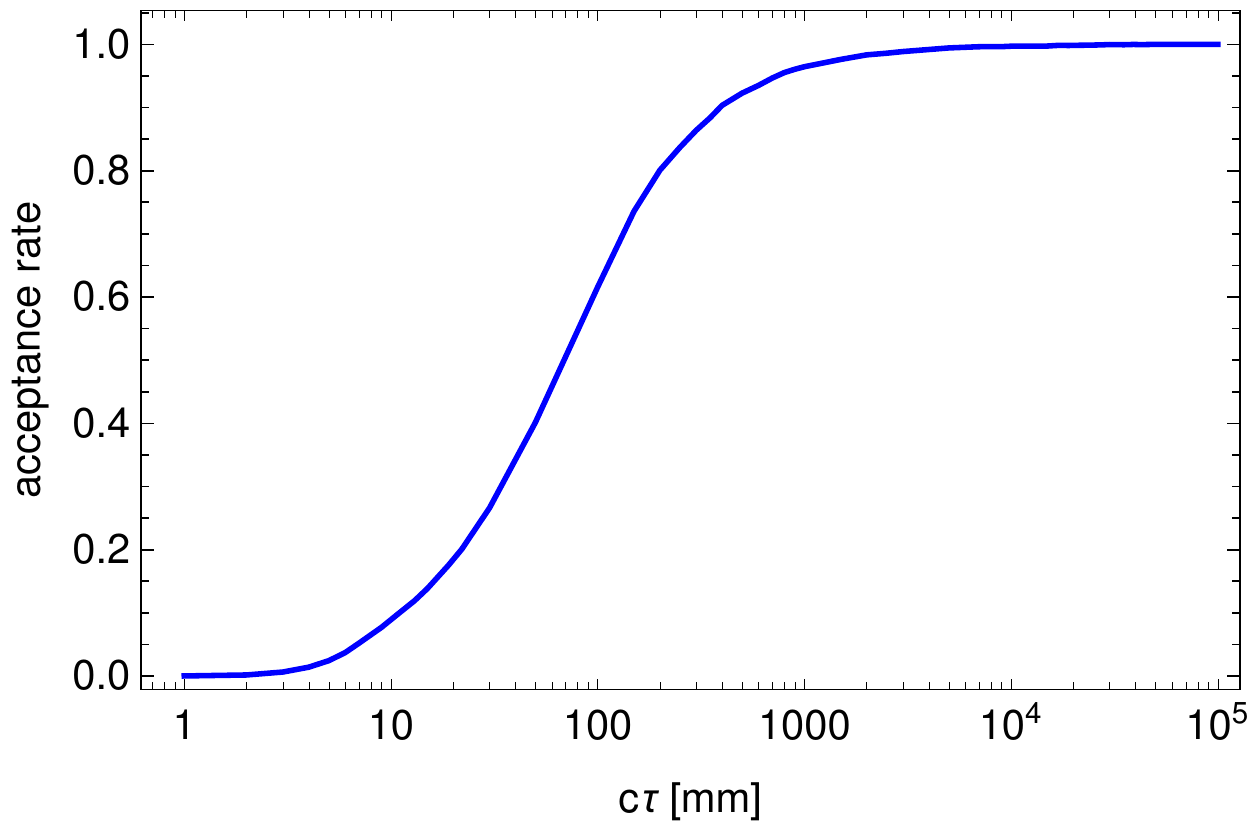}
       	\caption{MET search}
       	\label{acceptanceMET}
    \end{subfigure}
	\begin{subfigure}{0.47\textwidth}
		\centering
		\includegraphics[width=\linewidth]{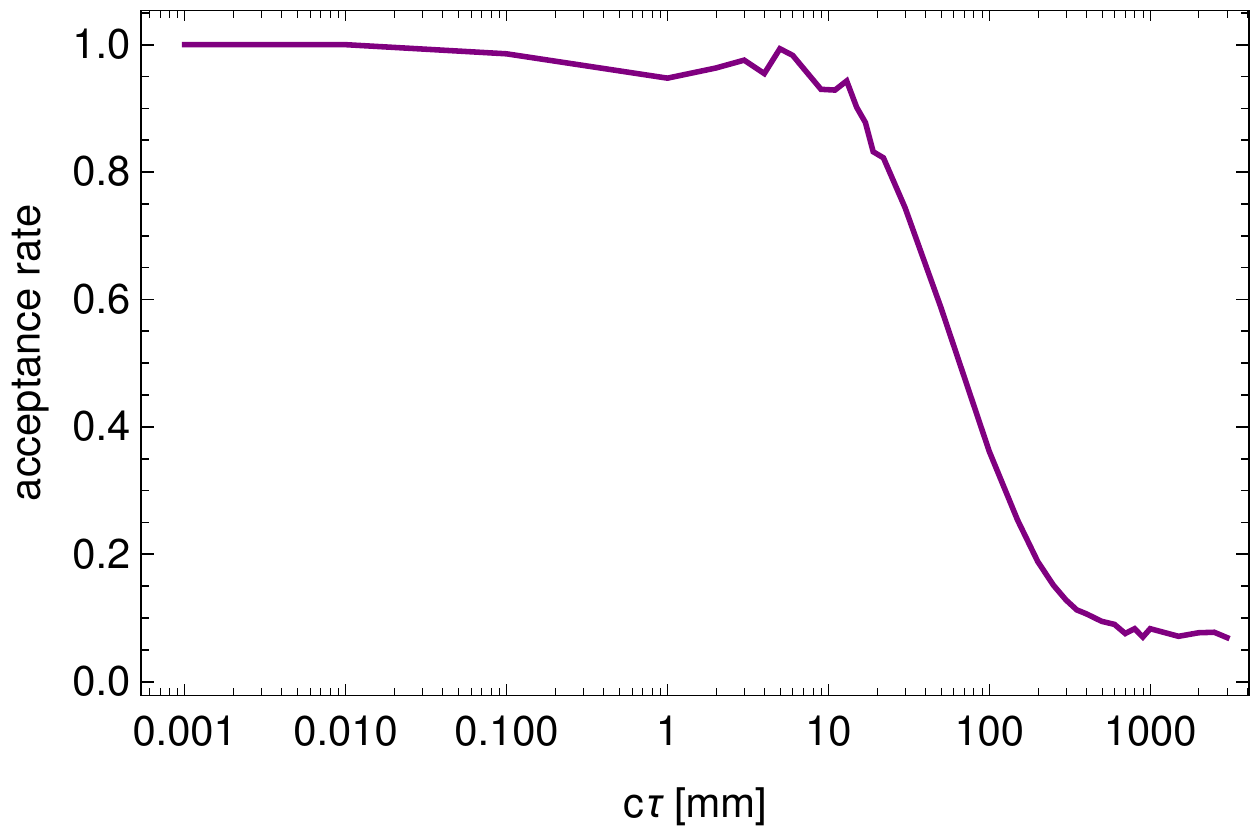}
		\caption{four jets search}
		\label{acceptance4Jets}
	\end{subfigure}
    \caption[Acceptance rate for the MET and 4 jets search]{Acceptance rate in the MET search (left) and four jets search (right), depending on the decay length $c\tau_{\pi_D}$.}
    \label{fig:acceptance}
\end{figure}

The acceptance rates are multiplied with the mediator pair production cross section for each value of $c\tau_{\pi_D}$ to get the effective cross section $\sigma_{eff}=\frac{N(c\tau_{\pi_D})}{N(c\tau_0)}\cdot\sigma$, where $c\tau_0$ is the baseline lifetime of the respective search. For each lifetime the values of $m_X$ with $\sigma_{eff} > \sigma_{LHC}$ can be excluded because the predicted cross section exceeds the observed limit from the LHC searches. In this way we obtain a constraint on the mediator mass which is displayed in~\cref{Constraints} (see also \cref{ConstraintsB,ConstraintsC} in the appendix for different dark sector benchmark points). The blue shaded areas are excluded from the recast of the MET search in \cite{MetNew}, while the purple shaded areas are excluded from the recast of the four jets search in \cite{4jets}.

\FloatBarrier
\subsection{Jet plus emerging jet search}
In the intermediate lifetime regime the dark pions decay at displaced vertices in the detector, so that jets emerge at various places in the detector and at various radial distances from the collision point. 

As this model has already been studied in \cite{FlavouredDarkSec,EmergingJets,DarkQCDScale},  an emerging jets search has already been executed  
at CMS in \cite{jetEmergingJet1,jetEmergingJet2} and the results of this work can be included directly.
In \cite{jetEmergingJet1,jetEmergingJet2} four new variables are defined to classify the properties of an emerging jet and then tested in a number of selection sets.
Each event is required to have at least four jets with $|\eta| < 2.0$ from which either two have to be tagged as emerging or one tagged as emerging and large MET. Furthermore, each event has to pass a threshold momentum on the scalar sum of all hadronic jets to be considered as signal.
The results of this search are visible in green in~\cref{compareFig,Constraints,ConstraintsB,ConstraintsC}.

Nonetheless, we also recast this search and compare the recasted results to the ones from \cite{jetEmergingJet1,jetEmergingJet2}. If the results agree we can adopt the procedure used for the recast here to the flavoured scenario. 
To recast the jets and emerging jets search the acceptance rate used in \cite{jetEmergingJet1} 
is multiplied with the cross section of the respective mass $m_X$ and the luminosity of the detector ${\cal L}=16.1~{\rm fb}^{-1}$ to get the expected signal. The expected background is given for each parameter point ($m_X$, $c\tau_{\pi_D}$) in \cite{jetEmergingJet1}. From this data the signal to background ratio $\frac{S}{\sqrt{S+B}}$ is calculated for every parameter point.
Every point with $\frac{S}{\sqrt{S+B}}\geq 2$ can be excluded as the signal would have been seen already.
\begin{figure}[h]
	\centering
	\includegraphics[width=0.7\textwidth]{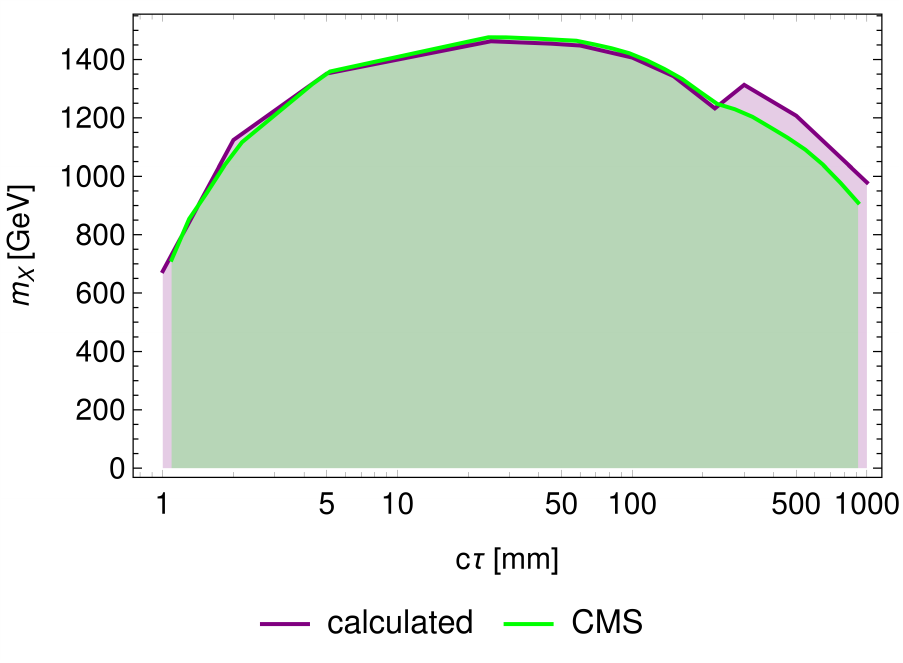}
	\caption[Constraints on the mediator mass from the emerging jets search]{Constraints on the mediator mass from the emerging jets search. In green: limit from CMS (Ref: \cite{jetEmergingJet1}). In purple: calculated limit from the simulated cross section; $m_{\pi_D}=5$~GeV.}
	\label{compareFig}
\end{figure}
\\The resulting bound on $m_X$ is displayed in~\cref{compareFig}. The calculated constraints (in purple) match the ones from CMS (in green) well.
Therefore, the procedure used here can be generalised safely to more than one flavour which is done in~\cref{sec:recast2}.

\FloatBarrier
\subsection{Summary of the collider searches}
The exclusion regions have been calculated according to the above described procedures for the three different benchmark points in~\cref{benchmarkTab}. As discussed in~\cref{sec:implementation}, these are the parameters set in the \textsc{Pythia8} Hidden Valley implementation, and do not necessarily agree with the Lagrangian parameters introduced in~\cref{sec:model}. In particular the ratios between the parameters should not be varied too much to ensure the proper functioning of the dark shower in \textsc{Pythia8}, and are kept fixed for our benchmark points. 
\begin{table}[h]
	\centering
	\begin{tabular}{|c|c|c|c|}
		\hline
		Parameter    & point A & point B & point C \\ \hline
		$m_{Q_D}$    & 10      & 4       & 20      \\ \hline
		$m_{\pi_D}$  & 5       & 2       & 10      \\ \hline
		$m_{\rho_D}$ & 20      & 8       & 40      \\ \hline
	\end{tabular}
	\caption{Summary of the different benchmark points.}
	\label{benchmarkTab}
\end{table} 

The results of these searches are summarised in~\cref{Constraints} for benchmark point A, as well as~\cref{ConstraintsB,ConstraintsC} in~\cref{appendixA} for benchmark points B and C. The coloured areas are the ones excluded by the respective searches.
\begin{figure}[h]
    \centering
    \includegraphics[width=0.7\textwidth]{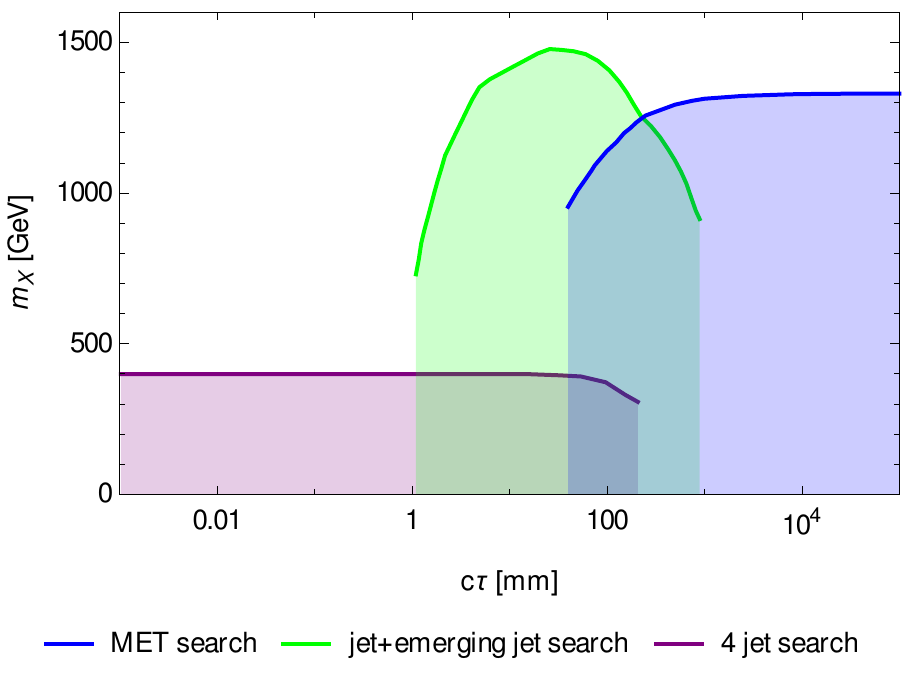}
    \caption[Constraints on the mediator mass for the three searches]{Constraints on the mediator mass for the three searches: (i) in purple: 4 jets search, (ii) in green: emerging jets search, (iii) in blue: MET search.}
    \label{Constraints}
\end{figure}

As expected, the searches cover different ranges of decay lengths and also vary in their overall sensitivity. The MET search covers lifetimes $c\tau \gtrsim 100$~mm. It constraints the respective parameter region up to $m_{X}\sim 1330$~GeV. The four jets search is less stringent. The bound on the mediator mass for lifetimes $c\tau\lesssim1$~mm is around $m_{X}\sim400$~GeV.

The emerging jets search is most efficient for decay lengths in the range of $c\tau_{\pi_D}\approx 1-100$~mm. In this range, mediator masses up to around $m_{X}\sim1500$ GeV can be excluded (in green). It should be noted here that this search uses less than 10\% of the now available data, while the MET limits already use almost all of the LHC run 2 data. We therefore expect the emerging jets search to increase its lead once it is updated. 

We find that for the most part the choice of benchmark point does not change the constraints of the mediator mass, with only minor differences appearing at the edges where the different searches loose their sensitivity, when comparing~\cref{Constraints} with ~\cref{ConstraintsB,ConstraintsC} in~\cref{appendixA}. Therefore, the limits on the mediator mass obtained here are robust and largely independent of the details of the dark sector.

\section{Searching for flavoured emerging jets}
\label{sec:recast2}
Next, we consider the case of not only one but several dark flavours.
To simulate events with multiple dark flavours, we use the function $\textit{HiddenValley:nFlav}$ in \textsc{Pythia8} as already implemented in the HV model.
\textsc{Pythia8} can only differentiate between flavour-diagonal and flavour-off-diagonal states. This means that we are only able to see two kinds of dark pions, but for different numbers of dark flavours they are weighted differently. For example, for $n_D = 2$ dark flavours there are $n_D = 2$ diagonal dark pions and $n^2_D-n_D-1 = 1$ off-diagonal dark pions. For $n_D = 3$ dark flavours, there are 2 diagonal dark pions and 6 off-diagonal dark pions. In the following, we let the lifetimes of the different dark pions vary independently, while their mass is set at $m_{\pi_D} = 5$~GeV (benchmark point A from ~\cref{benchmarkTab}). Each
of the searches from~\cref{sec:recast1} is redone with multiple combinations of dark pion lifetimes. 

The MET and four jet searches are recast as in~\cref{sec:recast1}, while a different strategy is needed for the emerging jets search. Here, the acceptance rates from \cite{jetEmergingJet1} are weighted according to the probability with which  diagonal and off-diagonal dark pions are produced in the hadronisation process. In \textsc{Pythia8} this is implemented using the string fragmentation picture, which results in the probability of finding a diagonal dark pion scaling as $1/n_D$. For the two and three flavour cases that we consider here, this implies either $1:1$ or $1:2$ ratios of diagonal to off-diagonal pions, and the corresponding acceptance rates. 

This new acceptance rate is then multiplied with the respective cross section and the integrated luminosity to get the expected signal for multiple combinations of lifetimes with the same mediator mass. We again take the expected background for every combination of lifetimes from \cite{jetEmergingJet1}. If the two lifetimes (at the same mediator mass) have different values for the expected background, the larger of the two is kept. This is a conservative approach which makes it less likely to exclude a region that should not be excluded. Again we exclude every point with $\frac{S}{\sqrt{S+B}}\gtrsim2$.

The results are shown in~\cref{fig:MET_flavoured} for the MET search,~\cref{fig:emerging_jet_flavoured} for the emerging jet search and in~\cref{fig:4jets_flavoured} for the four jets search. In all three figures the case of two flavours is shown in the left panel and the case of three flavours in the right panel.
\begin{figure}[!htb]
	\centering
	\begin{subfigure}{0.47\textwidth}
		\centering\includegraphics[width=\linewidth]{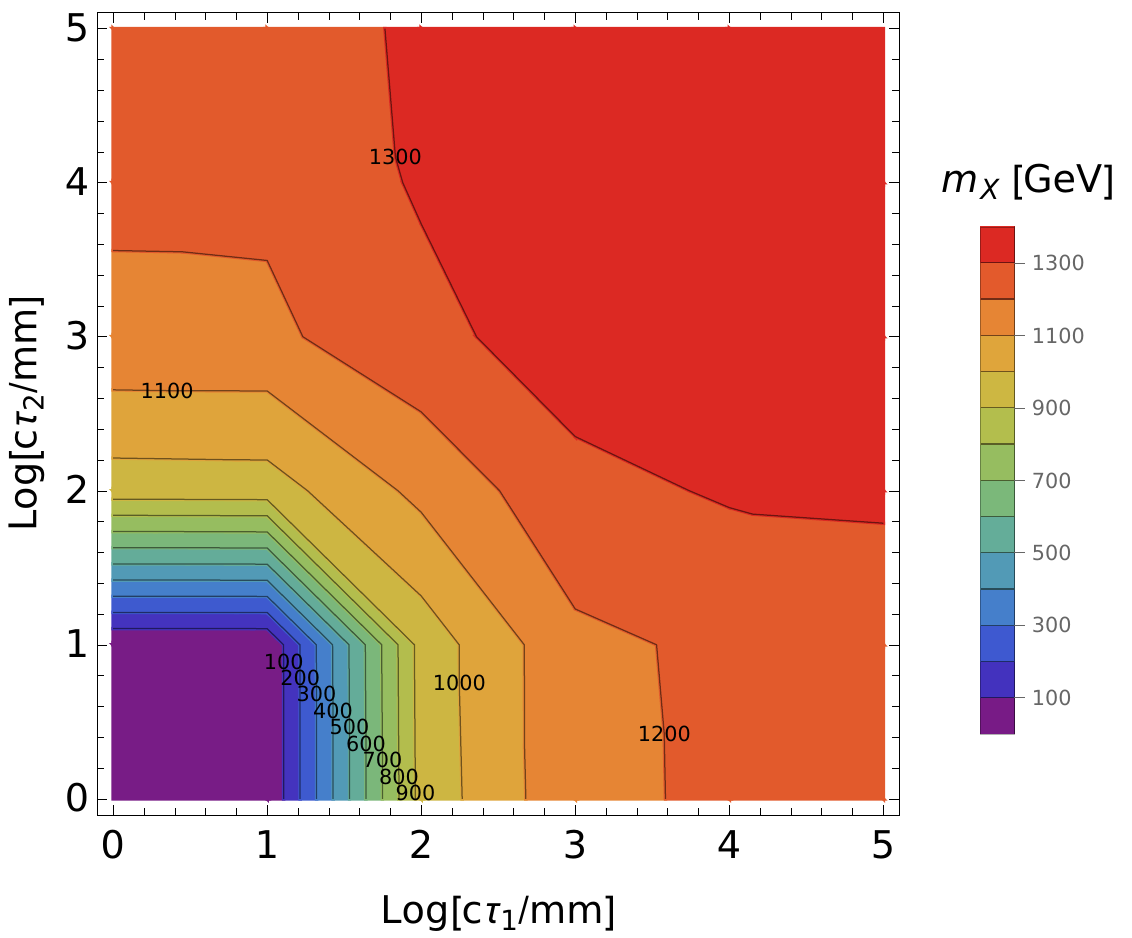}
		\caption{$n_D = 2$}
	\end{subfigure}
	\begin{subfigure}{0.47\textwidth}
		\centering\includegraphics[width=\linewidth]{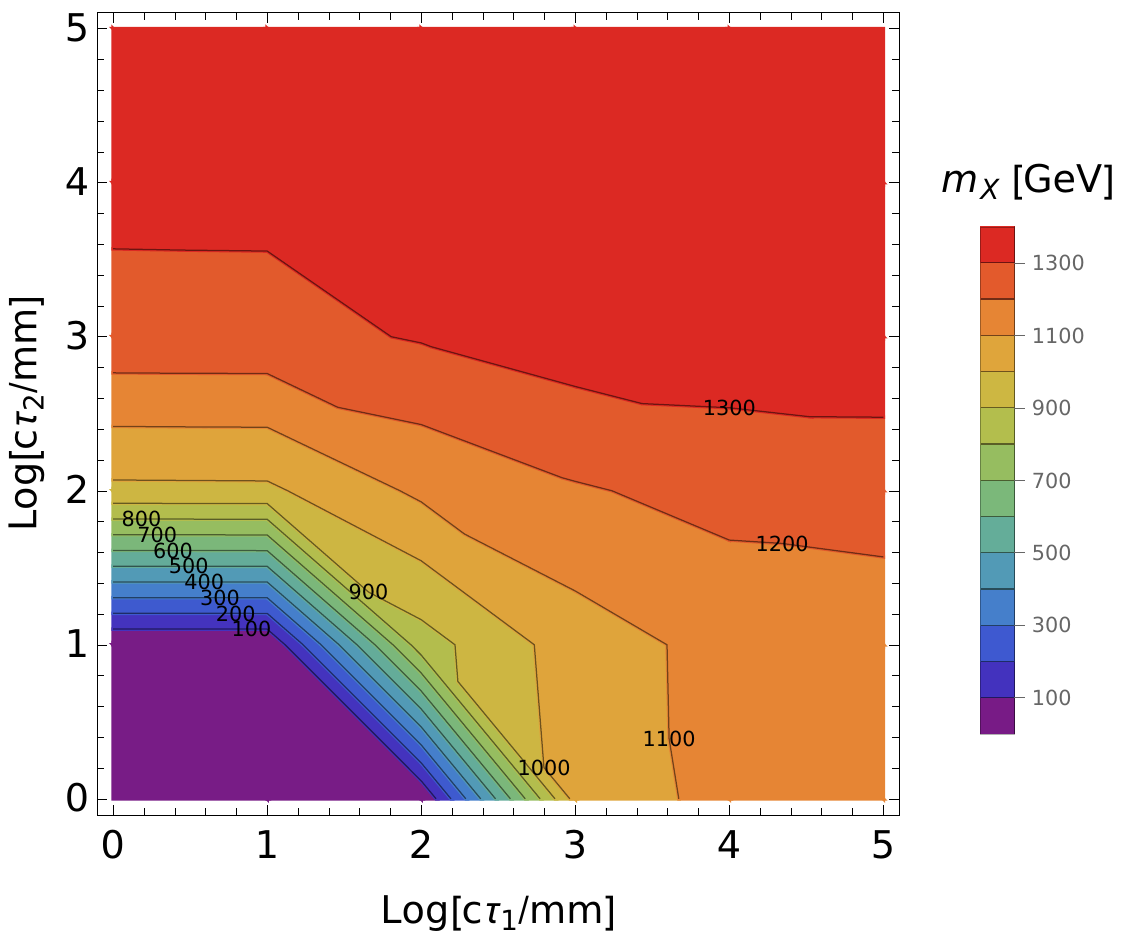}
		\caption{$n_D = 3$}
	\end{subfigure}
	\caption{Constraints on the mediator mass for two (left) and three  (right) dark flavours from the MET search with $c\tau_1$:
		diagonal dark pion decay length, $c\tau_2$: off-diagonal dark pion decay length. Values with $m_X$ smaller than the one displayed by the contour lines are excluded.}
	\label{fig:MET_flavoured}
\end{figure} 
\FloatBarrier
\begin{figure}[!htb]
	\centering
	\begin{subfigure}{0.47\textwidth}
		\centering\includegraphics[width=\linewidth]{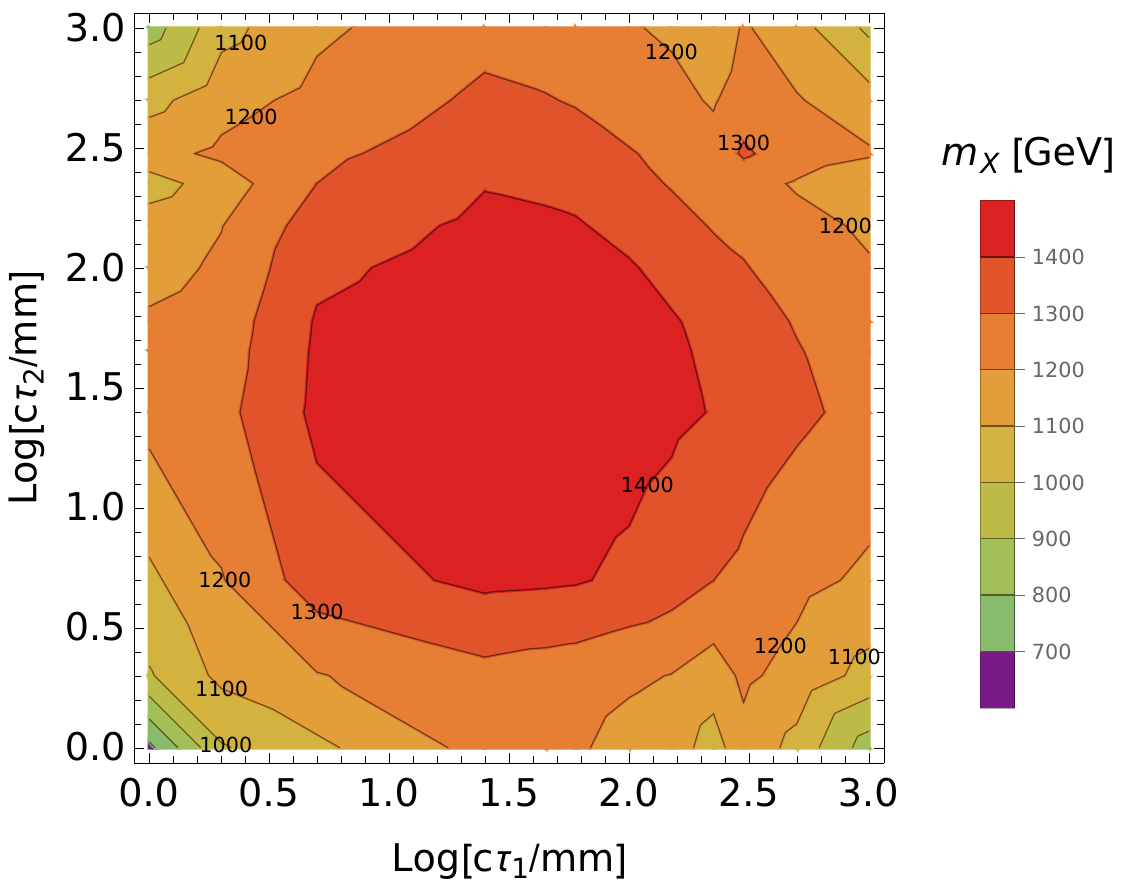}
		\caption{$n_D = 2$}
	\end{subfigure}
	\begin{subfigure}{0.47\textwidth}
		\centering\includegraphics[width=\linewidth]{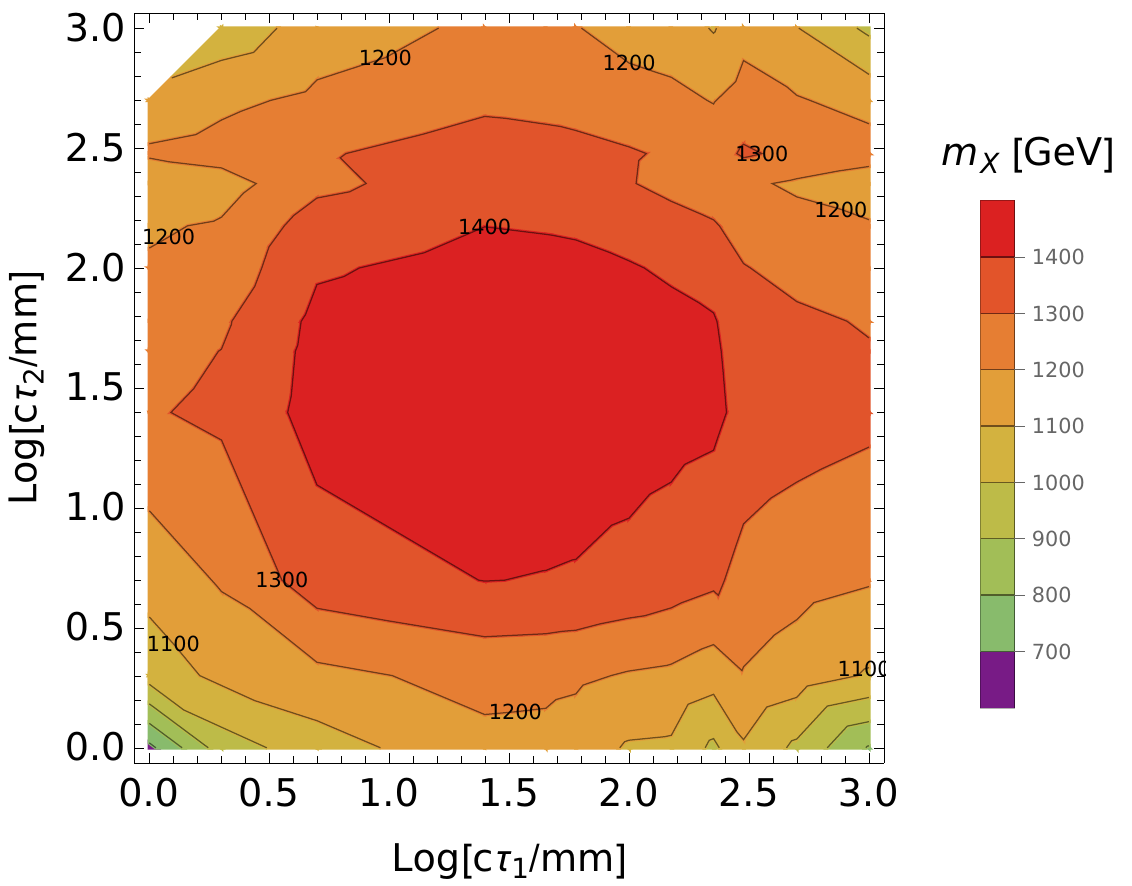}
		\caption{$n_D = 3$}
	\end{subfigure}
	\caption{Same as~\cref{fig:MET_flavoured}, but for the emerging jets search.}
	\label{fig:emerging_jet_flavoured}
\end{figure} 
\FloatBarrier
\begin{figure}[!htb]
	\centering
	\begin{subfigure}{0.47\textwidth}
		\centering\includegraphics[width=\linewidth]{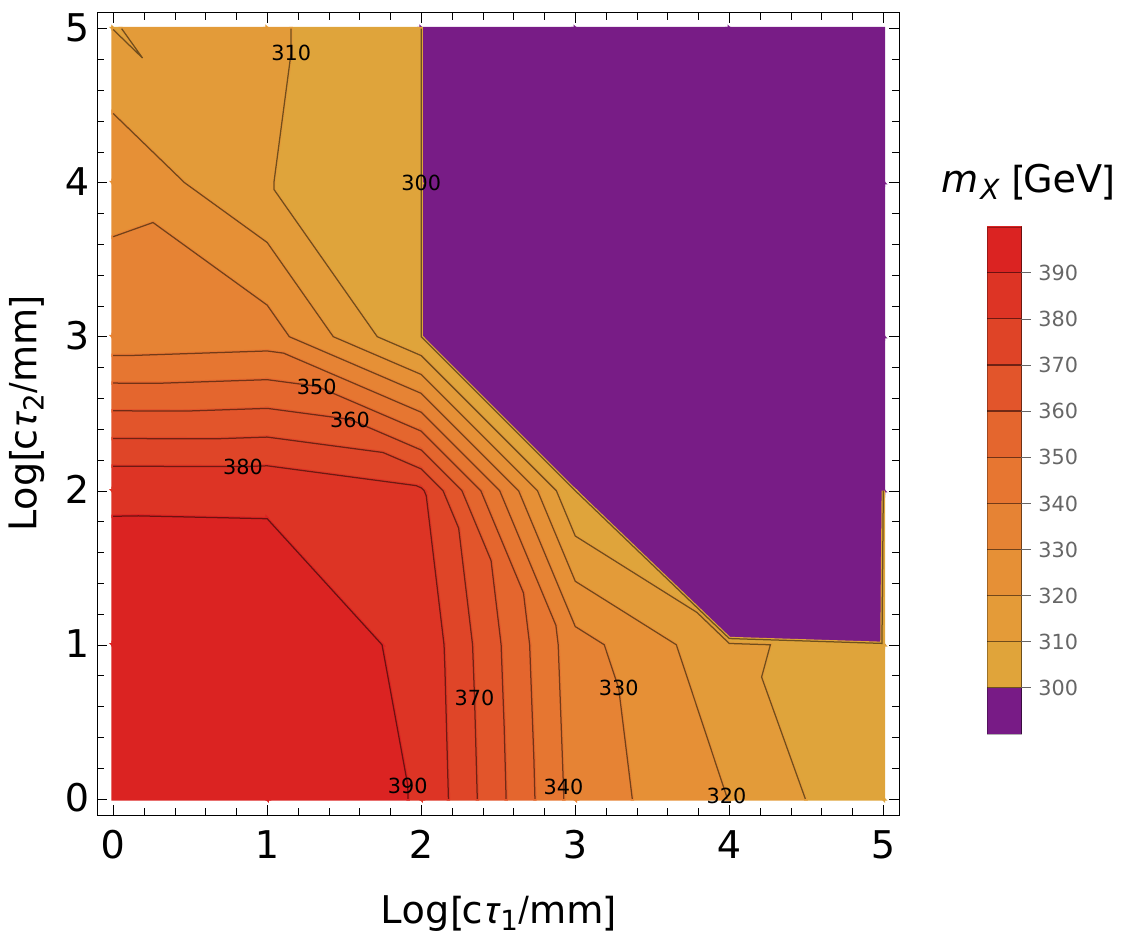}
		\caption{$n_D = 2$}
	\end{subfigure}
	\begin{subfigure}{0.47\textwidth}
		\centering\includegraphics[width=\linewidth]{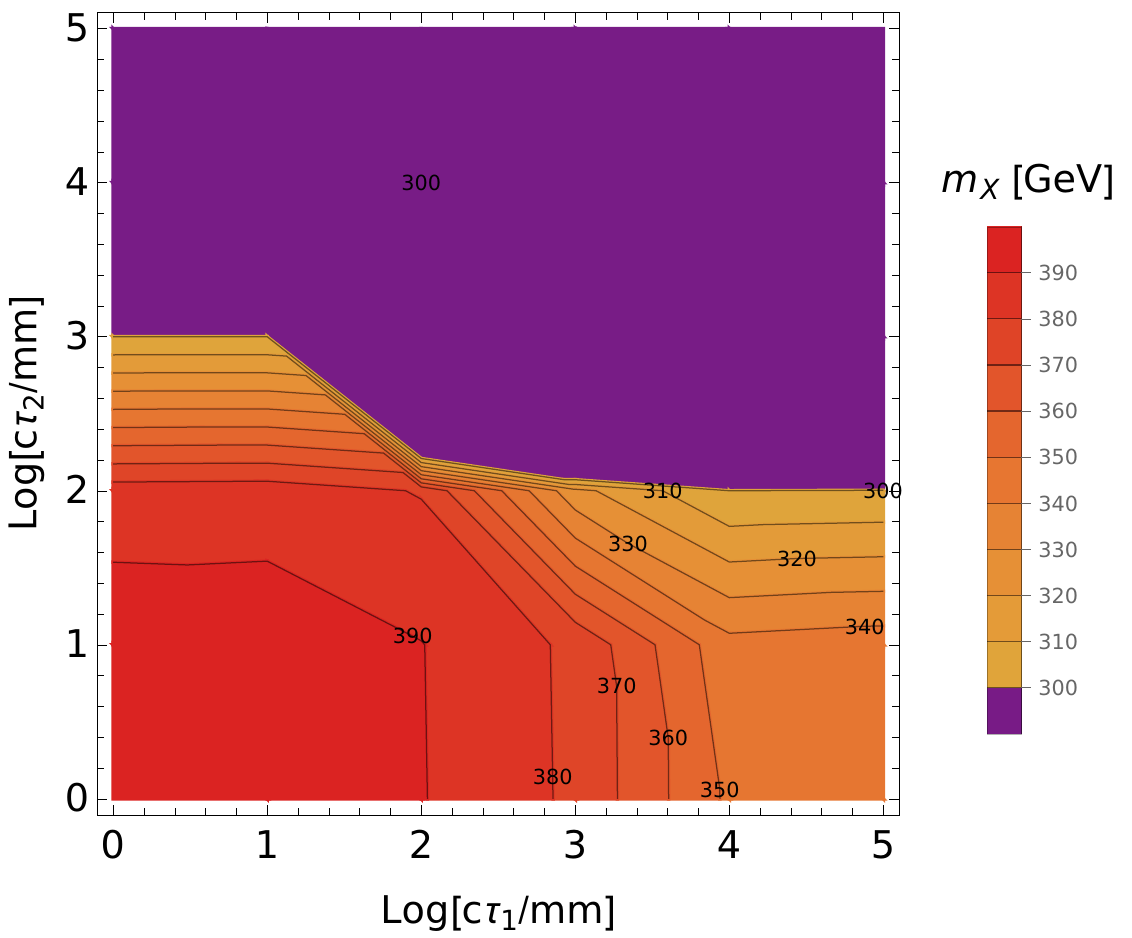}
		\caption{$n_D = 3$}
	\end{subfigure}
	\caption{Same as~\cref{fig:MET_flavoured}, but for the four jet search.}
	\label{fig:4jets_flavoured}
\end{figure}

The results are as one would expect. For $n_D=2$, the limits are symmetric in the lifetimes, and agree with the unflavoured result along the diagonal where both lifetimes are the same. When the number of flavours is increased, the limits depend more strongly on the lifetime of the off-diagonal dark pions. In particular the MET based search remains sensitive even if the diagonal dark pions decay promptly, because sufficient energy is carried away by the off-diagonal pions. Similarly, the four jets search retains some sensitivity if the diagonal dark pions decay mainly outside of the detector. 

Finally, in the emerging jets recast, we also observe the symmetric and slightly asymmetric limits in the $n_D =2$ and $n_D=3$ cases, respectively. Note that here we can only give limits for a smaller range of lifetimes for which the efficiencies are available. One can expect that this search will retain some sensitivity even if one of the dark pion lifetimes is sent to infinity, with moderately suppressed acceptance. On the other hand, if one of the dark pions becomes short lived, the effect on the acceptance will crucially depend on how the search is implemented experimentally. Obviously, if prompt or almost-prompt tracks are vetoed, as proposed in~\cite{EmergingJets}, the acceptance should drop to zero. Instead, other strategies which allow for some prompt activity might remain viable. This is certainly an interesting task for the experiments.

\section{Constraints on dark QCD}
\label{sec:constraints}
After finding the bounds from collider searches at the LHC, we will focus now on constraints from non-collider experiments. Additional constraints on dark QCD arise from flavour physics, direct detection experiments and big bang nucleosynthesis (BBN). These constraints are discussed in great detail in \cite{FlavouredDarkSec}. We will apply the constraints to the unflavoured as well as the flavoured case. In addition to combining collider and non-collider constraints on the model, we develop a prescription that allows us to cast the constraints in the dark matter mass-mediator mass plane. This allows for a better comparison of the different constraints, and corresponds to the scheme commonly adopted in collider searches for dark matter~\cite{Albert:2017onk}.

\subsection{Unflavoured scenario}
In the scenario with $\kappa_{\alpha i} = \mathbbm{1}\kappa_0$ no additional contributions to flavour changing processes are possible. Thus, here we only take into account the limits from BBN and direct detection experiments.

The dark pions can interfere with BBN as they can be long lived. To evade constraints from BBN at least one dark pion must have a lifetime smaller than one second as has been discussed in \cite{FlavouredDarkSec}. In the unflavoured case all dark pions have the same lifetime and therefore must have a lifetime $\tau< 1$~s.

The dark quark masses are degenerate and therefore they can form eight dark baryons. These can scatter off visible matter and be detected with dark matter direct detection experiments such as the XENON1T experiment. The matrix element of the dominant spin-independent scattering is \cite{Goodman:1984dc}
\begin{align}
\mathcal{M}_{p,n} =\sum_{a}\frac{|\kappa_{\alpha i}|^2}{8m_X^4}J_{D\alpha}^0J_{p,n}^0,
\end{align} 
with $J_{D\alpha}^0 = \sum_{k}\left\langle p_{D_k}|\bar{Q}_\alpha\gamma^0Q_\alpha|p_{D_k}\right\rangle$ and $J_{p,n}^0=\sum_{k}\left\langle p,n|\bar{d}\gamma^0d|p,n\right\rangle \approx 1,2$. The obtained averaged spin-independent cross section is \cite{FlavouredDarkSec}
\begin{align}
&\sigma_{N-D}^{SI} = \frac{1}{A}\sum_{a}\frac{(J_{Da}^0)^2|\kappa_{\alpha 1}^4\mu_{n-D}^2}{32\pi m_X^4}(J_n^0(A-Z)+J_p^0Z)^2\nonumber\\
&= \frac{1}{A}\sum_{a}\frac{|\kappa_{\alpha 1}^4\mu_{n-D}^2}{32\pi m_X^4}(2(A-Z)+Z)^2,
\end{align}
where $\mu_{n-D}$ is the reduced mass of the dark baryon-nucleus system and $Z = 54$ and $A = 131$ for xenon. We use the bounds from XENON1T \cite{Xenon} and the prospected bounds from DARWIN \cite{Darwin} to find which regions of the parameter space are still allowed.

Finally, we translate the bounds from the recasted collider searches discussed in~\cref{sec:recast1} from the mediator mass-dark pion lifetime plane to the dark matter-mediator mass plane using~\cref{lifetime}. With the assumption $f_{\pi_D} = m_{\pi_D}$~\cref{lifetime} connects the dark pion mass with the coupling matrix $\kappa$ and the lifetime of the dark pions. We choose three different scenarios for the diagonal coupling matrix $\kappa$: $\kappa_0 = 1$, $\kappa_0 = 0.1$ and the \textit{strange dark sector} scenario $\kappa_{11} = 0.01, \kappa_{22} = \kappa_{33} = 1$.  

\begin{figure}[t]
	\centering
	\begin{subfigure}{0.47\textwidth}
		\centering\includegraphics[width=\linewidth]{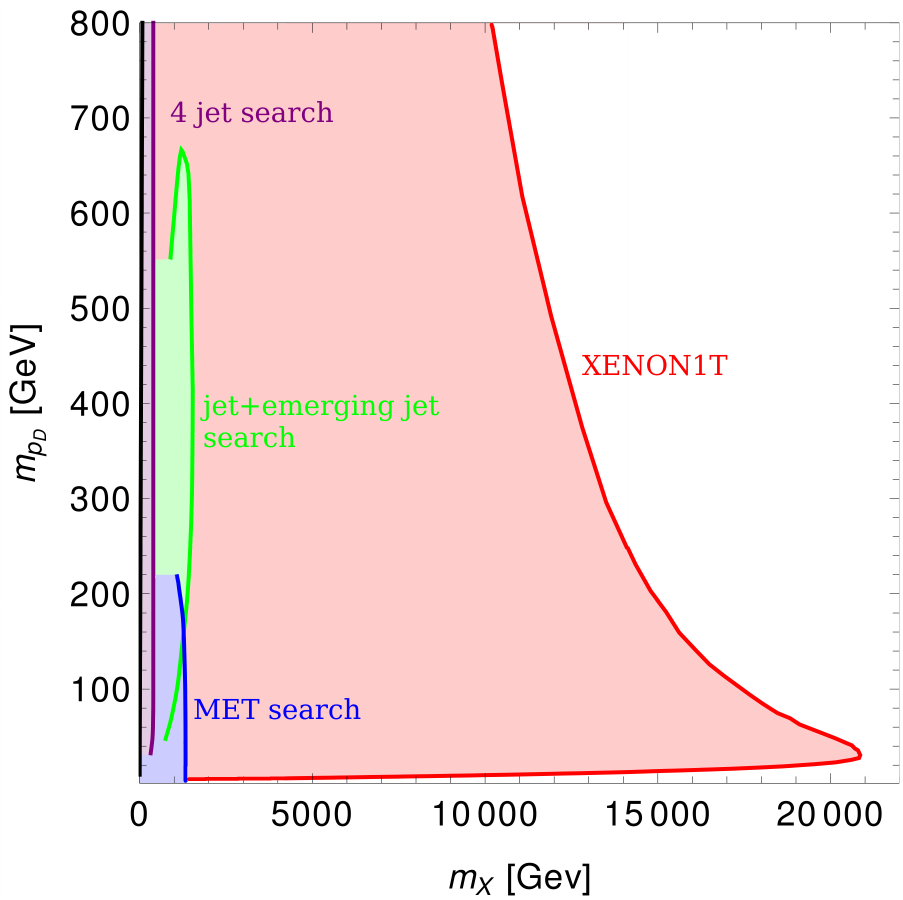}
		\caption{$\kappa_0 = 1$}
	\end{subfigure}
	\begin{subfigure}{0.47\textwidth}
		\centering\includegraphics[width=\linewidth]{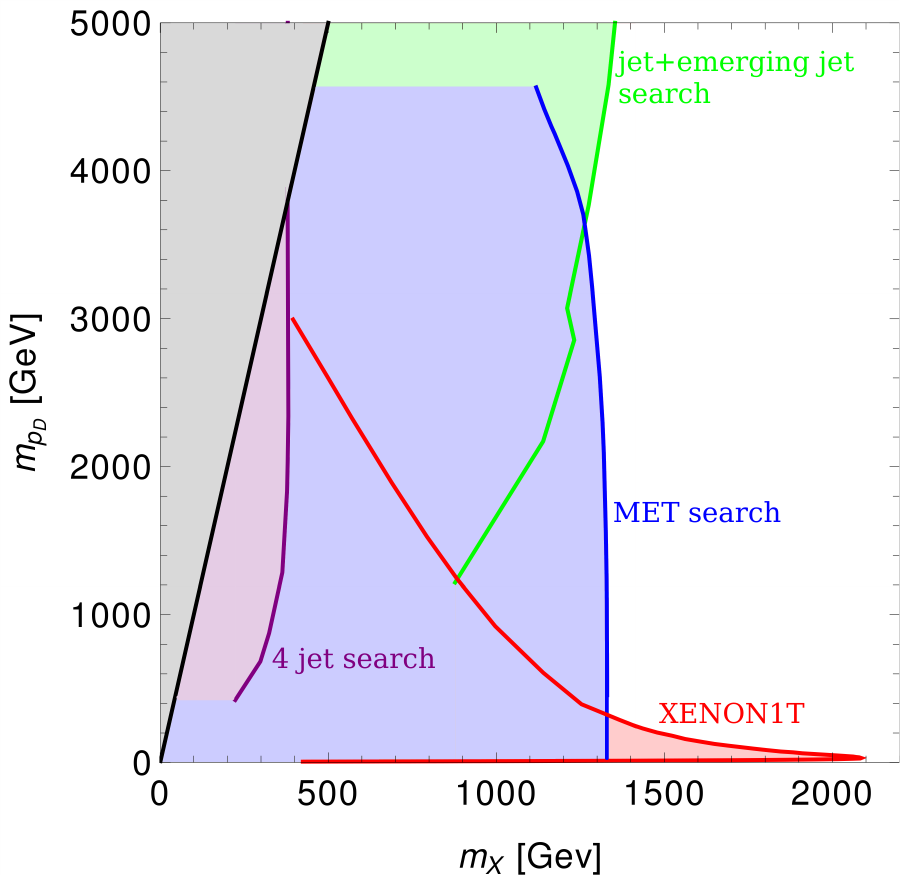}
		\caption{$\kappa_0 = 0.1$}
		\label{fig:b}
	\end{subfigure}
	\begin{subfigure}{0.47\textwidth}
		\centering\includegraphics[width=\linewidth]{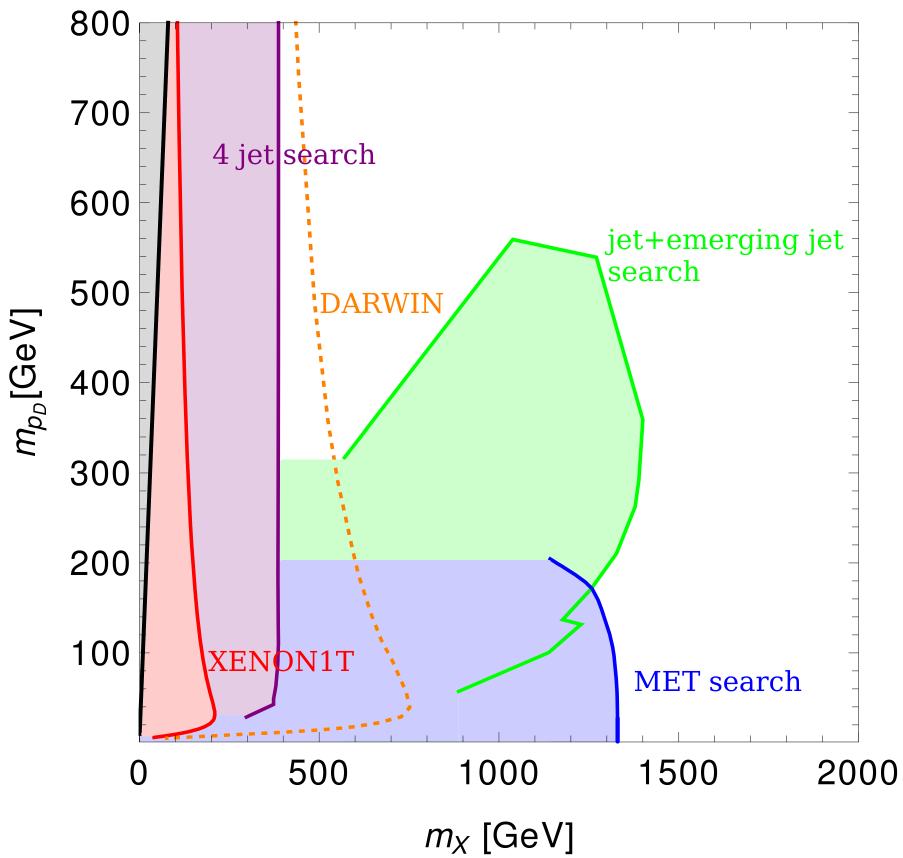}
		\caption{$\kappa_0 = 1$, $\kappa_{11} = \frac{\kappa_0}{100}$}
	\end{subfigure}
	\caption{Constraints on dark QCD in the $m_X-m_{p_D}$ plane for $m_{p_D} = 10m_{\pi_D}$. The green, blue and purple regions are excluded from the emerging jet, MET and four jet searches, the red region from XENON1T. The orange line shows the expected DARWIN exclusion limit. Above the black lines the dark pion mass is larger than the mediator mass.}
	\label{fig:all_constraints_diagonal}
\end{figure} 

For each scenario we can directly map~\cref{Constraints} to the dark pion mass-mediator mass plane. However, we want to consider the limits as a function of the mass of the dark matter candidate of the model, namely the lightest dark baryon. Therefore, we need to make assumptions about the relation between the mass of the dark matter candidate and the dark pions. We show the bounds for the case $m_{p_D} = 10 m_{\pi_D}$, which is based on the ratio of the proton to pion mass, in~\cref{fig:all_constraints_diagonal} and for $m_{p_D} = 3m_{\pi_D}$ in~\cref{fig:all_constraints_diagonal_3} in~\cref{appendixB}. 
In the shown parameter regions the dark pions have a lifetime $\tau< 1$~s, consequently BBN constraints are automatically satisfied.

From~\cref{fig:all_constraints_diagonal} it is obvious that for $\kappa_0 = 1$ direct detection gives the strongest bounds. This bound can be considerably relaxed for smaller values of $\kappa_0$. On the other hand, in the strange dark sector scenario where $\kappa_{11}$ is suppressed the emerging jet gives the most stringent constraint, while the direct detection bounds are strongly suppressed due to the small coupling to first generation quarks. Thus, the strange dark sector scenario has the best prospect to be discovered at colliders.

It is worth to notice that in the three scenarios different regimes of dark matter masses are displayed. This is due to the fact that the lifetime and mass of the dark pion are related with the fourth power of the coupling (c.f.~\cref{lifetime}). The grey shaded region shows the region where the dark pion mass would be larger than the mediator mass. It should also be noted that in~\cref{fig:b} the dark pion mass exceeds 100~GeV in some regions, so these limits should be treated with caution.

\subsection{Flavoured scenario}
In \cite{FlavouredDarkSec} the above limits as well as limits from flavour physics and discovery prospects at fixed target experiments were discussed in great detail for more complex flavour structures of $\kappa$. There, $\kappa$ is parametrised as $\kappa =VDU$, where $U$ is a $3\times 3$ unitary matrix, $V$ a $n_d\times n_d$ unitary matrix and $D$ a $n_d\times 3$ non-negative diagonal matrix. As before we choose $n_d = 3$. From the degeneracy of the dark quark masses a $U(n_d)_{dark}$ flavour symmetry allows to rotate $V$ away. $U$ can be decomposed into three unitary rotation matrices $U = U_{23}U_{13}U_{12}$ with 
\begin{align}
	U_{12} = \begin{pmatrix}
	\cos\theta_{12} & \sin\theta_{12}e^{-i\delta_{12}}&0\\
	-\sin\theta_{12}e^{-i\delta_{12}}&\cos\theta_{12}&0\\
	0&0&1
	\end{pmatrix}
\end{align} 
and $U_{13}$ and $U_{23}$ analogously, where $\theta_{ij}$ are the new mixing angles and $\delta_{ij}$ CP-phases. In the following we use $\delta_{ij}=0$. Finally, $D$ can be written as
\begin{align}
	D = \Big(\kappa_0\cdot\mathbbm{1} + diag(\kappa_1,\kappa_2,-(\kappa_1+\kappa_2))\Big)
\end{align}  \cite{Agrawal:2014aoa}.
With this parametrisation we now have also off-diagonal terms in the coupling matrix. Thus, we also need to take flavour changing processes like neutral meson mixing and flavour changing decays of kaons and B mesons into account.\\
It was found in \cite{FlavouredDarkSec} that for neutral meson mixing the dependence of $\theta_{ij}$ drops out if $D_{ii}$ and $D_{jj}$ are degenerate. Consequently, the limits from neutral meson-mixing can not only be evaded if all angles are small but also if only one is large and the corresponding entries in $D$ are nearly degenerate. This leads to three useful scenarios:
\begin{align}
&\boldsymbol{ij = 12:} \text{ }\kappa_0 = 1, \text{ }\kappa_1=\kappa_1,\text{ }\kappa_2=0,\text{ }\theta_{12}=\theta_{12},\text{ }\theta_{13}=0,\text{ }\theta_{23}=0, \\
&\boldsymbol{ij = 13:} \text{ }\kappa_0 = 1, \text{ }\kappa_1=\kappa_1,\text{ }\kappa_2=0,\text{ }\theta_{12}=0,\text{  } \theta_{13}=\theta_{13},\text{ }\theta_{23} =0, \\
&\boldsymbol{ij = 23:} \text{ }\kappa_0 = 1, \text{ }\kappa_1=\kappa_1,\text{ }\kappa_2=0,\text{ }\theta_{12}=0,\text{ }\theta_{13}=0,\text{ }\theta_{23}=\theta_{23} .
\end{align}
For the three scenarios all bounds are calculated in \cite{FlavouredDarkSec} for a mediator mass $m_X = 1$~TeV. As we have seen above this mediator mass is already excluded in the region where the emerging jet and jets plus MET searches are sensitive. Consequently, we rescale the calculated bounds for a mediator mass $m_X = 1.55$~TeV, which is allowed in all three search regions. Hereby, we use the fact that all considered processes constraining the parameter space are proportional to $\left(\frac{\kappa_0}{m_X}\right)^4$.

\begin{figure}[t]
	\centering
	\begin{subfigure}{0.47\textwidth}
		\centering\includegraphics[width=\linewidth]{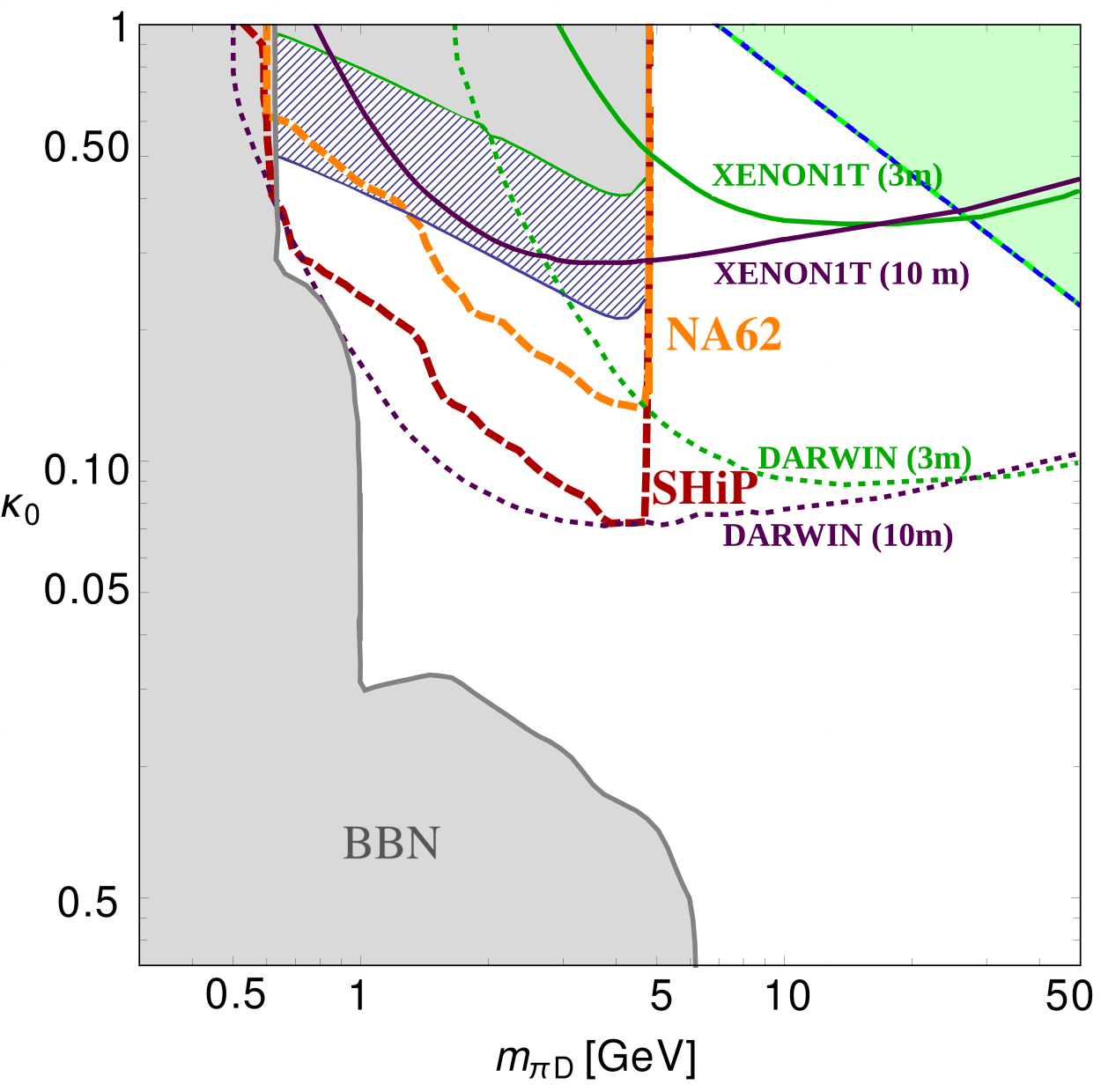}
		\caption{"13 scenario"}
	\end{subfigure}
	\begin{subfigure}{0.47\textwidth}
		\centering\includegraphics[width=\linewidth]{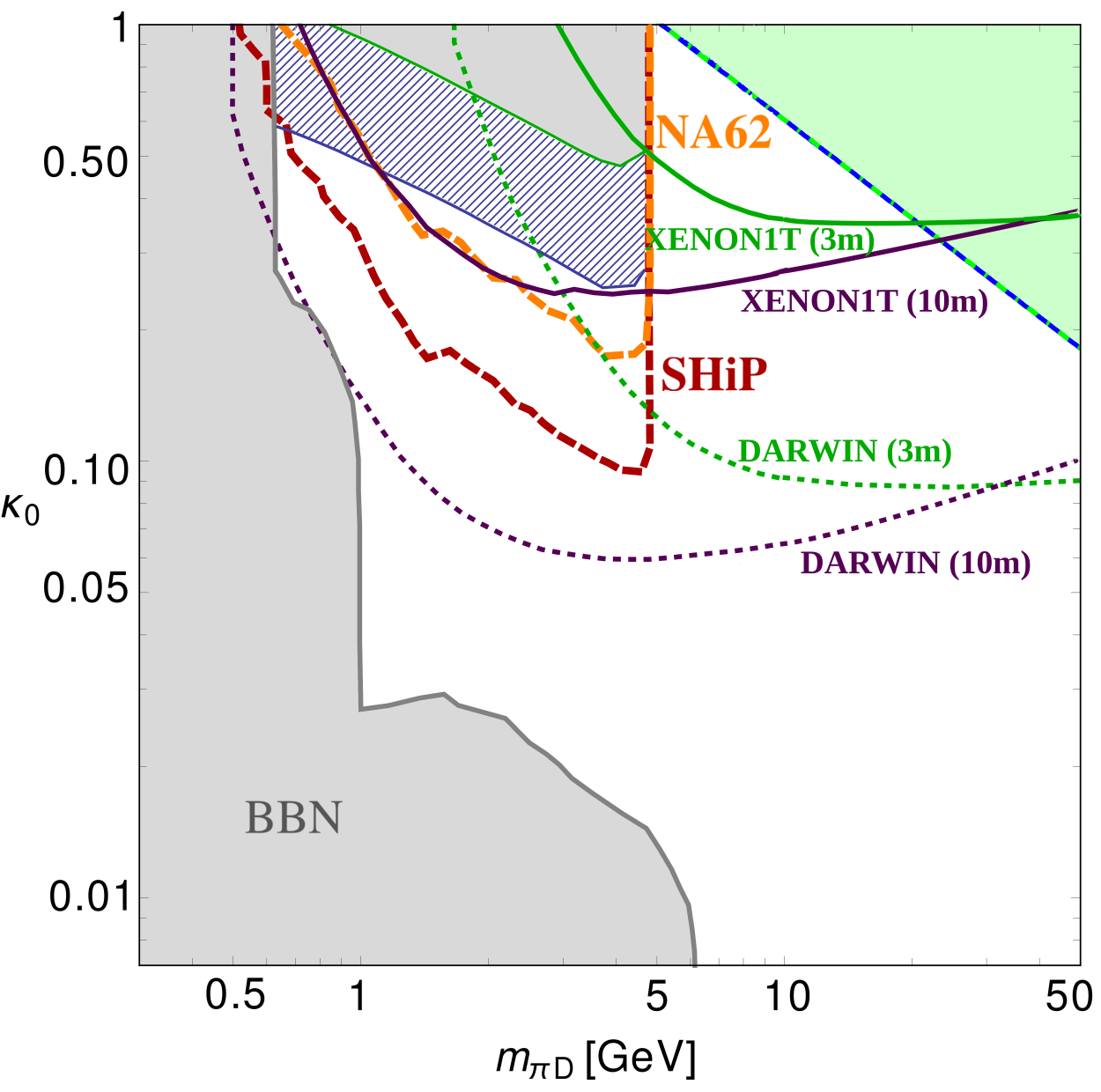}
		\caption{"23 scenario"}
	\end{subfigure}
	\caption{Constraints on dark QCD in the $m_{p_D}-\kappa_0$-plane for the "13" and "23" scenario. Above the dark red (orange) line more than 3 dark pion decays are expected over the full run in SHiP (NA62 in beam-dump mode). In green ($m_{p_D} = 3m_{\pi_D}$) and dark purple ($m_{p_D} = 10m_{\pi_D}$) are current (solid) and projected (dashed) dark matter direct detection bounds. The grey region is disfavoured from BBN or flavour constraints. In the blue hatched region the upcoming measurements of Belle II for $Br(B\to K\bar{\nu}\nu)$ is expected to be sensitive~\cite{Belle1,Belle2,Belle3}. The region above the blue-green line will be probed by future emerging jets searches, while below the MET search should be most sensitive. The bounds have been rescaled for $m_X = 1.55$~TeV and $f_D = m_{\pi_D}$.}
	\label{fig:mpid_k0}
\end{figure}

Furthermore, the LHC bounds calculated in~\cref{sec:recast1} can also be applied to the non-diagonal coupling matrix scenarios as we only allow for small off-diagonal elements and have already seen in~\cref{sec:recast2} that the overall results of the searches are the same for flavoured scenarios. In~\cref{fig:mpid_k0} the bounds that were obtained in \cite{FlavouredDarkSec}  for the  "13" (left panel) and "23" scenario  (right panel) are shown together with the regions in which the discussed LHC searches will be sensitive for a dark mediator mass $m_X = 1.55$~TeV.   

The light green shaded regions in~\cref{fig:mpid_k0} show the region in which the dark pion lifetime falls in the range of sensitivity of the emerging jets search. Below the green and blue line the MET search should eventually probe the model. Note that with the chosen mediator mass the three  discussed collider searches do not exclude any of the parameter space yet. 

Overall, one can see that in both the "13" and the "23" scenario the MET search will be most relevant for the parameter space under consideration. Only for higher values of both $\kappa_0$ and $m_{\pi_D}$ the emerging jet search becomes more sensitive, with part of that region already excluded by XENON1T.

\section{Summary}
\label{sec:summary}

In this work, constraints on dark QCD from collider searches, flavour physics and cosmology were studied and updated. For the first time constraints on the mediator mass are calculated for the full dark meson lifetime range for an unflavoured scenario. 
The results are summarised in~\cref{Constraints,ConstraintsB,ConstraintsC}.
This excludes mediator masses to up to 1.5 TeV for the emerging jets search and masses up to around 400 GeV and 1.3 TeV for the four jets and the two jets + MET search, respectively. One can also see that the shape of the constraints is independent from the dark pion mass.

Furthermore, we also presented these results in the usual mediator mass - dark matter mass plane and analysed the constraints arising from BBN as well as direct detection experiments. The main results for the unflavoured case are shown for different values of $\kappa_0$ and different dark pion - dark baryon mass ratios in~\cref{fig:all_constraints_diagonal} and~\cref{fig:all_constraints_diagonal_3}. With $\kappa_0 = 1$ the direct detection bounds are ruling out large parts of the parameter space. This can be avoided by either choosing $\kappa_0$ smaller or introducing a \textit{strange dark sector}. For smaller $\kappa_0$ the direct detection bound is still the strongest constraint for small dark baryon masses and the collider searches rule out dark matter masses $m_{p_D}\sim\mathcal{O}(1000\text{GeV})$. In the strange dark sector the coupling to down quarks is suppressed compared to the coupling to strange and bottom quarks. Therefore, scattering of dark baryons on protons and neutrons is suppressed, while the production of dark pions at hadron colliders is not. Then, the direct detection bound is weaker and the collider searches give the strongest constraints on the parameter space. Still, a large part of the parameter space is not yet probed in such a scenario. With a strange dark sector collider experiments at LHC are the best opportunities to discover dark QCD. 

The unflavoured analysis can be generalised to multiple flavours, c.f.~\cref{sec:recast2}. By recasting the collider searches for the case of two and three flavours we found that the general form of the constraints is the same. With this in mind, the constraints from direct detection experiments, flavour physics and BBN are considered and applied to the model. In~\cref{fig:mpid_k0} the results for the flavoured cases are depicted and combined with the constraints arising from flavour physics. Small dark pion masses are ruled out by BBN.  For $m_{\pi_D} < m_B-m_K$ large couplings can be probed with Belle and Belle II measurements of $Br(B\to K\nu\bar{\nu})$. Fixed target experiments like NA62 and FASER can probe to lower couplings in the same mass region. In the higher dark pion mass region XENON1T rules out large couplings, the exact area depending on the dark matter mass to dark pion mass ratio. Part of the still allowed parameter space can be probed via the emerging jet search for high mediator masses.  However, in most of the remaining, unexplored parameter space the dark pions are longer lived, and MET searches should provide the best opportunities to further probe that region at colliders. 

For the scenarios with two or more dark pion lifetimes, more tailored search strategies could further improve the sensitivity. One relatively simple option would be to require some amount of missing energy along with the emerging jets search. Furthermore, one could focus more on searching for displaced subjets inside multi-jet signatures, or require reconstructed vertices in the muon system in addition to prompt or displaced signatures at shorter distances from the interaction point. Due to the nature of the dark shower, these different signatures are still expected to be aligned along the dark jets, while most existing searches attempt to isolate them radially. Therefore, we believe that an interesting challenge awaits to be solved here.

\section*{Acknowledgements}
We thank A.~Belloni, S.~Eno and E.~Bernreuther for useful discussions and encouragement.
Work in Mainz was supported by the Cluster of Excellence Precision Physics, Fundamental Interactions, and Structure of Matter (PRISMA+ EXC 2118/1) funded by the German Research Foundation(DFG) within the German Excellence Strategy (Project ID 39083149), and by grant 05H18UMCA1 of the German Federal Ministry for Education and Research (BMBF).
HM is supported by the German Research Foundation DFG through the RTG 2497 and the CRC/Transregio 257.

\appendix

\section{Benchmark point B and C}
For completeness, in this appendix we show the constraints on the mediator mass for different dark sector benchmark points. It can be seen that the individual constraints have some mild dependence on the dark pion masses, however once taken together the overall constraint on the mediator mass remains almost unchanged. 
\label{appendixA}
\begin{figure}[h]
	\centering
	\includegraphics[width=0.7\textwidth]{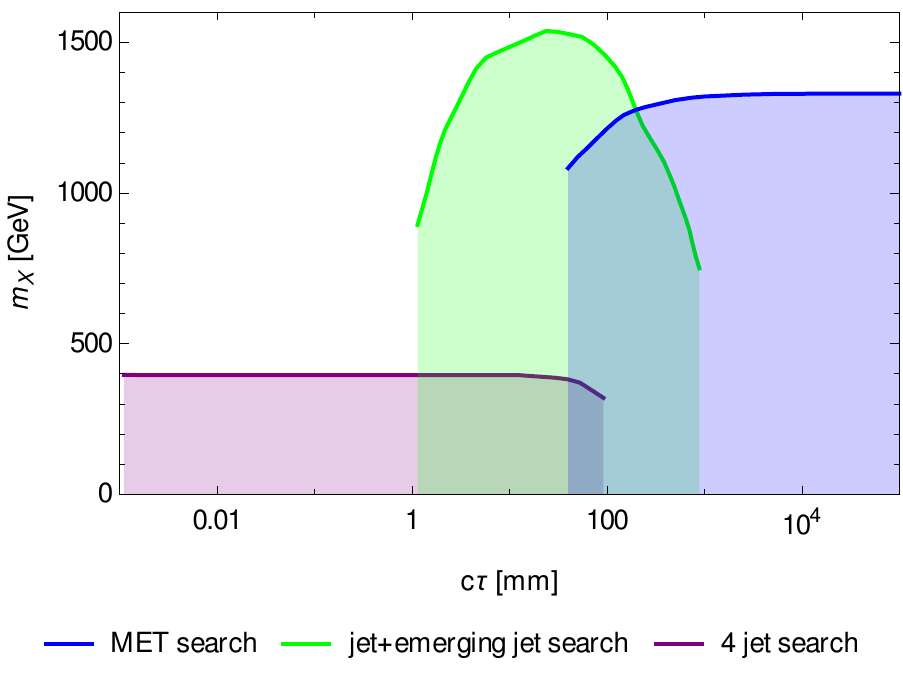}  
	\caption[Constraints on the mediator mass for benchmark point B for the three searches]{Constraints on the mediator mass for benchmark point B for the three searches: (i) in purple: 4 jets search, (ii) in green: emerging jets search, (iii) in blue: MET search.}
	\label{ConstraintsB}
\end{figure}
\begin{figure}[h]
	\centering
	\includegraphics[width=0.7\textwidth]{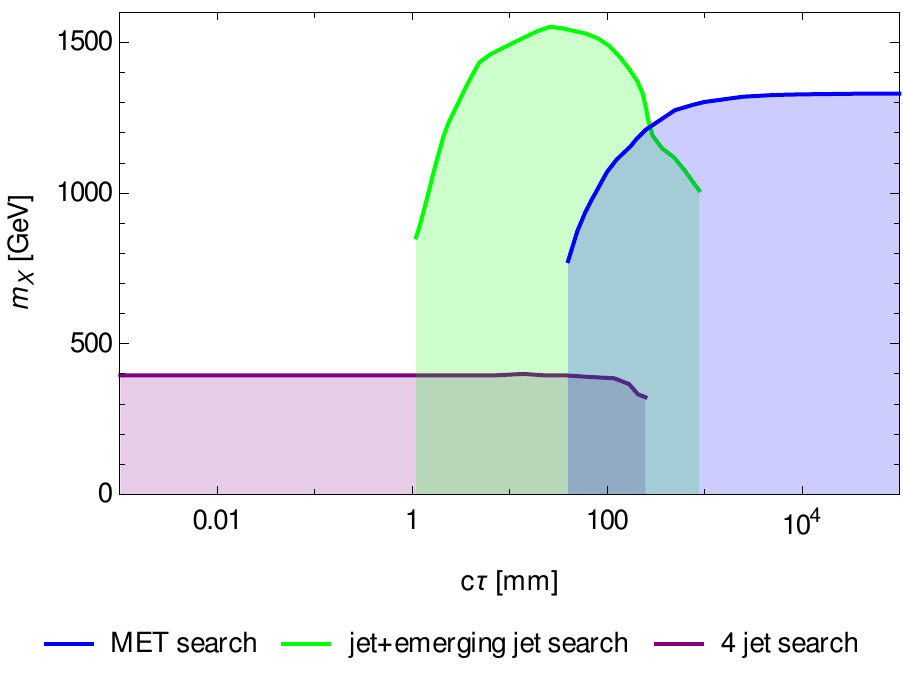}
	\caption[Constraints on the mediator mass for benchmark point C for the three searches]{Constraints on the mediator mass for benchmark point C for the three searches: (i) in purple: 4 jets search, (ii) in green: emerging jets search, (iii) in blue: MET search.}
	\label{ConstraintsC}
\end{figure}

\newpage
\section{Constraints on dark QCD for different DM dark pion mass ratios}
\label{appendixB}
In~\cref{sec:constraints}, we showed the combined constraints of collider and non-collider searches on the dark QCD model, for a fixed ratio of dark baryon to dark pion mass, namely $m_{p_D} = 10 m_{\pi_D}$. Since the mass of the baryons is tied to the confinement scale, while the pion mass is essentially a free parameter, we show here in addition the limits for $m_{p_D} = 3 m_{\pi_D}$ instead. The direct detection constraint (red shaded region) only depends on the DM and mediator mass, and is therefore unchanged. We have seen above that variations in the dark pion mass between 2~GeV and 10~GeV also affect the collider limits only mildly. However, it should be noted that in some regions in these figures, the dark pion mass exceeds 100~GeV, so these limits should be treated with caution. 

\begin{figure}[!htb]
	\centering
	\begin{subfigure}{0.47\textwidth}
		\centering\includegraphics[width=\linewidth]{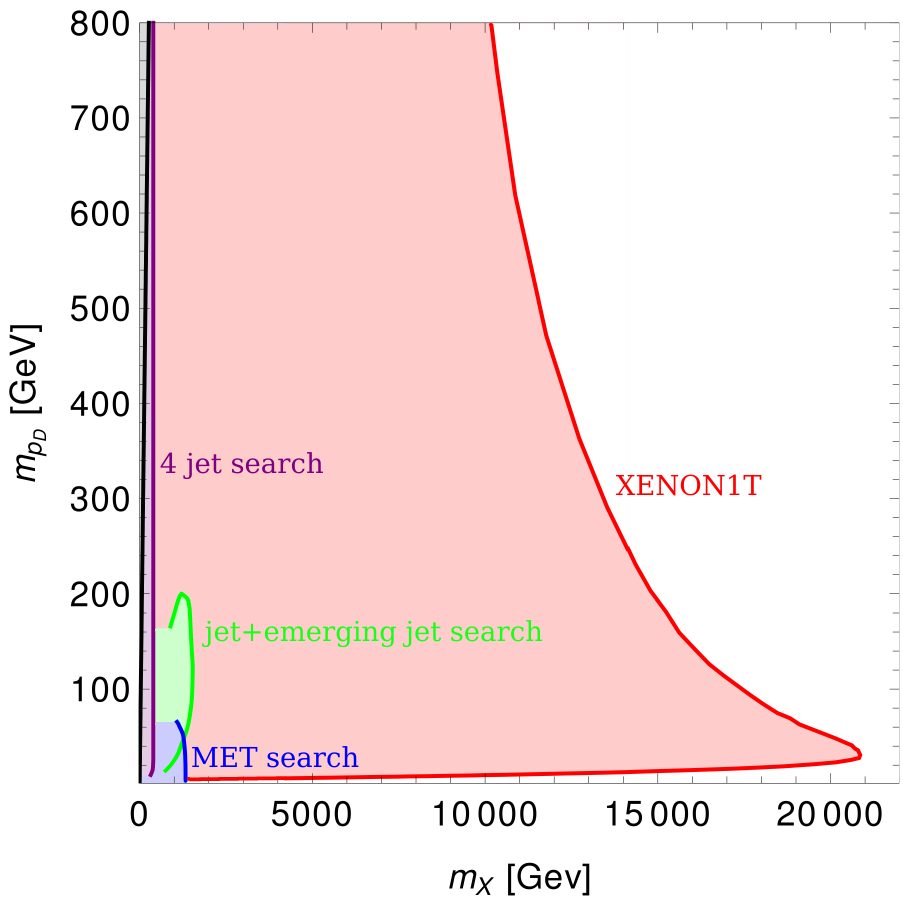}
		\caption{$\kappa_0 = 1$}
	\end{subfigure}
	\begin{subfigure}{0.47\textwidth}
		\centering\includegraphics[width=\linewidth]{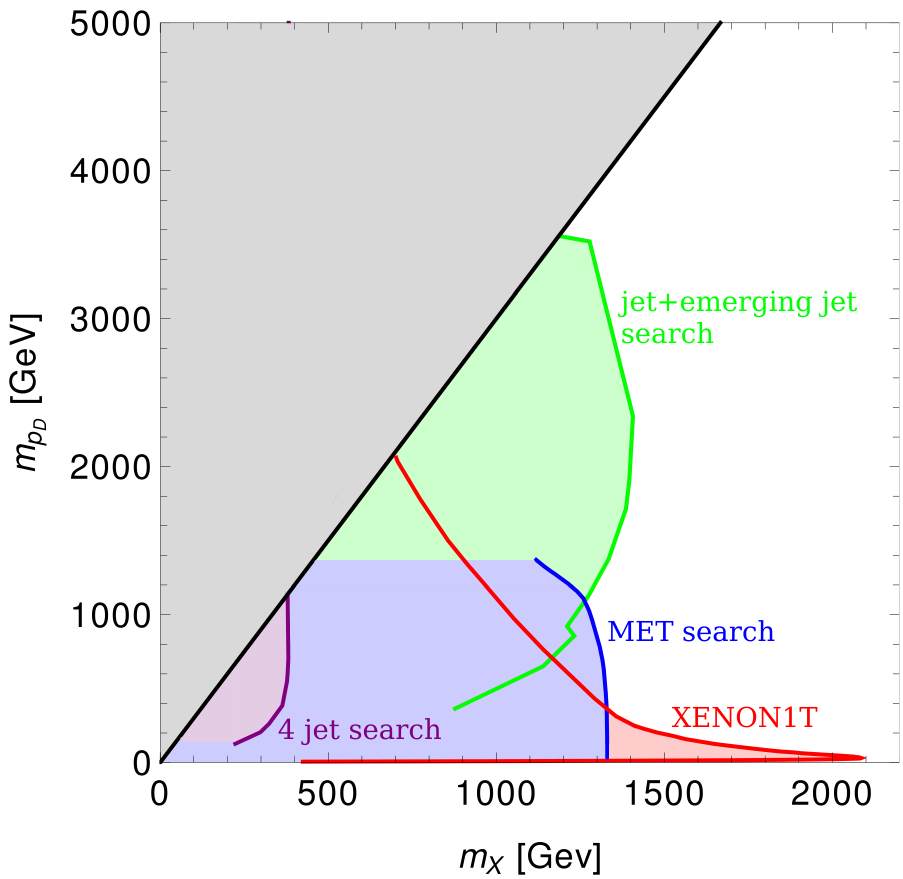}
		\caption{$\kappa_0 = 0.1$}
	\end{subfigure}
	\begin{subfigure}{0.47\textwidth}
		\centering\includegraphics[width=\linewidth]{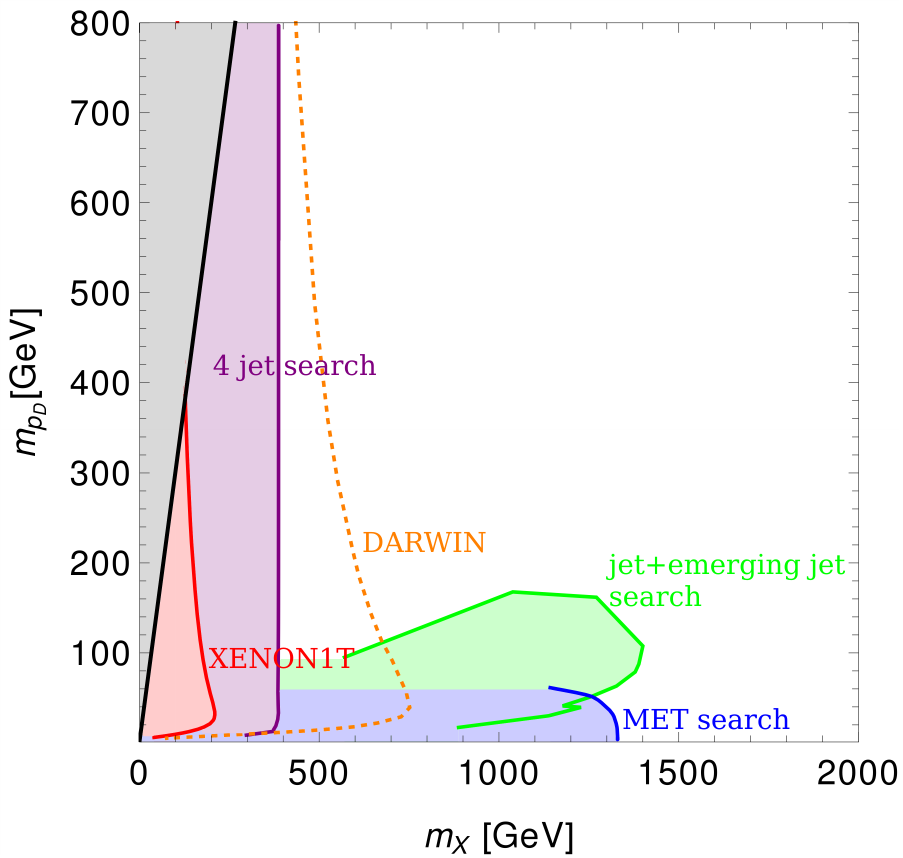}
		\caption{$\kappa_0 = 1$, $\kappa_{11} = \frac{\kappa_0}{100}$}
	\end{subfigure}
	\caption{Constraints on dark QCD in the $m_X-m_{p_D}$-plane for $m_{p_D} = 3m_{\pi_D}$. The green, blue and purple regions are excluded from the emerging jet, MET and four jet searches. The red region from XENON1T. The orange line shows the expected DARWIN exclusion limit. Above the black lines the dark pion mass is larger than the mediator mass.}
	\label{fig:all_constraints_diagonal_3}
\end{figure} 
\FloatBarrier

\bibliographystyle{JHEP}
\bibliography{dark}

\end{document}